\documentclass[fleqn,usenatbib]{mnras}

\usepackage{fix-cm}

\usepackage{newtxtext,newtxmath}

\usepackage[T1]{fontenc}

\DeclareRobustCommand{\VAN}[3]{#2}
\let\VANthebibliography\thebibliography
\def\thebibliography{\DeclareRobustCommand{\VAN}[3]{##3}\VANthebibliography}

\usepackage{graphicx}	
\usepackage{amsmath}	

\usepackage{ulem}

\usepackage{multirow}
\usepackage{threeparttable}
\usepackage{threeparttablex}
\usepackage{longtable}

\newcommand{\uas}  {$\mu$as}

\definecolor{myred}{RGB}{255,66,56}

\title[VLBI astrometry of 11 radio stars]{VLBI astrometry of radio stars to link radio and optical celestial reference frames - II. 11 radio stars}

\author[Zhang et al.]{Jingdong Zhang,$^{1,2,3}$
Bo Zhang,$^{1}$\thanks{E-mail: zb@shao.ac.cn}
Shuangjing Xu,$^{4,1}$
Xiaofeng Mai,$^{1,2,5}$
Mark J. Reid,$^{6}$
Pengfei Jiang,$^{7}$\newauthor
Wen Chen,$^{8,9}$
Fengchun Shu,$^{1}$
Jinling Li,$^{1}$
Lang Cui,$^{7}$
Xingwu Zheng,$^{10}$
Yan Sun,$^{1}$
and Zhaoxiang Qi$^{1,2}$
\\
$^{1}$Shanghai Astronomical Observatory, Chinese Academy of Sciences, 80 Nandan Road, Shanghai 200030, People’s Republic of China\\
$^{2}$University of Chinese Academy of Sciences, No.19 (A) Yuquan Rd, Shijingshan, Beijing 100049, People’s Republic of China\\
$^{3}$Department of Geodesy and Geodynamics, Finnish Geospatial Research Institute (FGI), National Land Survey of Finland, Vuorimiehentie 5,\\Espoo 02150, Finland\\
$^{4}$Korea Astronomy and Space Science Institute, 776 Daedeok-daero, Yuseong-gu, Daejeon 34055, Republic of Korea\\
$^{5}$University of Helsinki, P.O. Box 64, FI-00014, Finland\\
$^{6}$Harvard-Smithsonian Center for Astrophysics, 60 Garden Street, Cambridge, MA 02138, USA\\
$^{7}$Xinjiang Astronomical Observatory, Chinese Academy of Sciences, 150 Science 1-Street, Urumqi 830011, People’s Republic of China\\
$^{8}$Yunnan Observatories, Chinese Academy of Sciences, Kunming 650216, Yunnan, People’s Republic of China\\
$^{9}$Yunnan Key Laboratory of the Solar Physics and Space Science, Kunming 650216, People’s Republic of China\\
$^{10}$School of Astronomy and Space Science, Key Laboratory of Modern Astronomy and Astrophysics (Ministry of Education), Nanjing University, \\Nanjing 210023, People’s Republic of China\\
}

\date{Accepted XXX. Received YYY; in original form ZZZ}

\pubyear{2025}

\begin{document}
\label{firstpage}
\pagerange{\pageref{firstpage}--\pageref{lastpage}}
\maketitle

\begin{abstract}
    The alignment between the radio-based International Celestial Reference Frame (ICRF) and the optical \textit{Gaia} Celestial Reference Frame (\textit{Gaia}-CRF) is critical for multi-waveband astronomy, yet systematic offsets at the optical bright end ($G \lesssim 13$) limit their consistency.
    While radio stars offer a potential link between these frames, their utility has been restricted by the scarcity of precise Very Long Baseline Interferometry (VLBI) astrometry.
    In this study, we present new VLBI astrometry of 11 radio stars using the Very Long Baseline Array (VLBA), expanding the existing sample with positions, parallaxes, and proper motions measured.
    All 11 radio stars were detected, for 10 of which parallaxes and proper motions can be estimated, achieving median uncertainties better than 0.1\,mas and 0.1\,mas yr$^{-1}$, respectively.
    These new samples greatly contribute to the link between ICRF and \textit{Gaia}-CRF at the optical bright end.
\end{abstract}




\section{Introduction}
\label{sect:intro}

The establishment of a unified celestial reference frame (CRF) covering all bands is critical for advancing multi-waveband astronomy, geodesy, and deep-space navigation.
At radio wavelengths, the International Celestial Reference Frame \citep[ICRF,][]{1998AJ....116..516M} is the radio realization of the International Celestial Reference System \citep[ICRS,][]{1995A&A...303..604A} and is defined through Very Long Baseline Interferometry (VLBI) observations of extragalactic sources (quasars). Its latest iteration, ICRF3, achieved a coordinate noise floor of 30\,\uas\ \citep{2020A&A...644A.159C}.
In the optical band, the third data release of the \textit{Gaia} mission \citep[\textit{Gaia} DR3,][]{2023A&A...674A...1G} provides astrometry for over 1.6 million quasars, defining the optical realization of the ICRS, the \textit{Gaia} Celestial Reference Frame 3 (\textit{Gaia}-CRF3), which has comparable precision to ICRF3 \citep{2022A&A...667A.148G}.

\textit{Gaia}-CRF3 is aligned to ICRF3 at a 7\,\uas\ level (formal uncertainty) using common quasars \citep{2022A&A...667A.148G}.
However, systematic discrepancies persist for optical bright sources ($G \lesssim 13$) in \textit{Gaia} DR3, attributed to magnitude-, color-, and galactic-latitude-dependent shifts in centroid determinations and instrument calibration limitations, such as the Window Class (WC) effects \citep{2020A&A...633A...1L}.
These systematics manifest as orientation offsets and residual spins between the bright \textit{Gaia}-CRF3 and the quasar-based ICRF3 \citep{2018A&A...616A...2L,2018ApJS..239...31B}, necessitating independent validation to achieve an accurate and stable alignment between the CRFs.

Given the optical faintness of most quasars, efforts to link the frames rely on radio stars, which are observable in both radio and optical (bright-end) bands \citep{2016MNRAS.461.1937M}.
Several studies have attempted to align \textit{Gaia}-CRF to ICRF with radio stars, e.g., \citet{2020A&A...633A...1L,2022AstL...48..790B,2023A&A...676A..11L,2024A&A...689A.134L,2024MNRAS.529.2062Z}.
However, the small number of available radio stars restricts the robustness of the alignment: positions and proper motions of only several tens of radio stars have been measured with VLBI.

It is also possible to assess the bright \textit{Gaia}-CRF internally within \textit{Gaia}, which is not restricted by sample numbers.
For example, \citet{2021A&A...649A.124C} estimated spin parameters with samples collected from known binaries and open clusters.
However, orientation parameters cannot be estimated through this method.

It is therefore crucial to increase the number of radio stars with VLBI measurements; to this end, we have undertaken observations within the framework of a long-term research initiative.
Observing strategies for radio star astrometry with VLBI are discussed in \citet{2024MNRAS.529.2062Z}.
Our pilot Very Long Baseline Array (VLBA) program (obs. ID: BZ077) successfully measured parallaxes and proper motion of two radio stars, HD 199178 and AR Lac \citep{2023MNRAS.524.5357C}.
In another program (VLBA obs. ID: BZ080), we observed three radio stars: HD 179094, RZ Cas, and SZ Psc (Jiang et al. \textit{in prep.}).
This study presents new VLBI astrometry for 11 radio stars, focusing on the data reduction and astrometric results, while the frame link results with all the new data are presented in \citet{2025A&A...699A.345Z}.

The paper is organized as follows: Sect. \ref{sect:obs} describes the VLBI observations; Sect. \ref{sect:reduce} details the data reduction; Astrometric results are presented in Sect. \ref{sect:result}; Finally we summarize in Sect. \ref{sect:summary}.

\section{Observations}
\label{sect:obs}

Our observations were conducted under VLBA programs BZ087, BZ103, and BZ107.
The target selection is introduced in Sect. \ref{sect:target}, while the observation setups and scheduling are introduced in Sect. \ref{sect:setup} and \ref{sect:schedule}, respectively.

\subsection{Target selection}
\label{sect:target}

We selected target radio stars from the Very Large Array (VLA) historical observations.
For example, 46 radio stars were observed with the Very Large Array Plus Pie Town \citep{2007AJ....133..906B}.
These stars are detectable on baselines of tens of kilometers; however, it is in doubt whether they are bright enough to be detected on baselines of thousands of kilometers.
Therefore, we confirmed through snapshot observation whether they can be detected with VLBI.
The snapshot was conducted under the European VLBI Network (EVN) program EZ029 at X band, with 11 antennas (Wb, Ef, Mc, O6, T6, Ur, Ys, Ir, Sv, Zc, Bd), recording rate of 1 Gbps, and on-source time of about five minutes for each star.
Six stars (HD 199178, IM Peg, SZ Psc, RZ Cas, RS CVn, and V1762 Cyg) were detected with an SNR\,$>$\,5, in which three (HD 199178, SZ Psc, and RZ Cas) were observed in previous programs (obs. ID: BZ077 and BZ080), and one was selected to be observed in this program (RS CVn).

The Very Large Array Sky Survey \citep[VLASS,][]{2021ApJS..255...30G} provides a great number of candidates for VLBI observation.
Through crossmatching the VLASS catalog with the SIMBAD \citep{simbad} database and \textit{Gaia} DR2 \citep{2018A&A...616A...1G}, we selected 73 radio star candidates that are point-like and have a brightness of $>$\,1 mJy/beam.
Then we confirmed their detectability through another snapshot observation (EVN obs. ID: EX009) at C band.
A recording rate of 2 Gbps, and an on-source time of about ten minutes for each star were adopted, providing twice the sensitivity as EZ029.
20 candidates were detected, and we selected ten to be observed in this program.
Information of all 11 selected stars is listed in Table \ref{table:target}, which all have no parallax and proper motion measured with VLBI before.
The stars are divided into five groups according to their right ascension for easier scheduling.

\begin{table*}
    \centering
    \setlength{\tabcolsep}{4pt}
    \caption{Optical information of selected radio stars}
    \label{table:target}
    \begin{threeparttable}
        \begin{tabular}{clcccccrrrrc}
            \hline
            Group &\makebox[30pt][c]{Name} & Type & RA (J2000) & DEC (J2000) & $\sigma_{\alpha*}$ & $\sigma_{\delta}$ & \makebox[46pt][c]{$\varpi$} & \makebox[46pt][c]{$\mu_{\alpha*}$} & \makebox[46pt][c]{$\mu_{\delta}$} & \makebox[12pt][c]{$G$} & RUWE \\
            & & & ($^{\mathrm{h}}$ $^{\mathrm{m}}$ $^{\mathrm{s}}$) & ($^{\mathrm{\circ}}$ $^{\mathrm{\prime}}$ $^{\mathrm{\prime\prime}}$) & (mas) & (mas) & \makebox[46pt][c]{(mas)} & \makebox[46pt][c]{(mas yr$^{-1}$)} & \makebox[46pt][c]{(mas yr$^{-1}$)} & \makebox[12pt][c]{(mag)} & \\
            \hline
            \multirow{2}*{A} & FF Aqr & EB & 22:00:36.4256 & $-$02:44:26.869 & 0.030 & 0.030 & 4.734$\pm$0.034 & 31.284$\pm$0.037 & -10.173$\pm$0.035 & 9.1 & 2.1 \\
            & HD 8357 & XB & 01:22:56.7569 & $+$07:25:09.332 & 0.038 & 0.028 & 21.981$\pm$0.045 & 94.296$\pm$0.048 & 231.124$\pm$0.034 & 7.1 & 1.6 \\
            \hline
            \multirow{2}*{B} & EI Eri & RS CVn & 04:09:40.8924 & $-$07:53:34.176 &  0.030 & 0.022 & 18.380$\pm$0.039 & 37.847$\pm$0.039 & 102.570$\pm$0.032 & 7.0 & 1.9 \\
            & V1859 Ori & T Tau & 05:22:54.7927 & $+$08:58:04.679 & 0.014 & 0.011 & 3.480$\pm$0.016 & 1.375$\pm$0.016 & -9.264$\pm$0.012 & 9.9 & 1.1 \\
            \hline
            \multirow{2}*{C} & V1355 Ori & RS CVn & 06:02:40.3604 & $-$00:51:37.225 & 0.018 & 0.016 & 7.842$\pm$0.023 & 11.859$\pm$0.024 & 11.693$\pm$0.020 & 8.9 & 1.2 \\
            & AR Mon & RS CVn & 07:20:48.4543 & $-$05:15:35.797 & 0.021 & 0.022 & 2.298$\pm$0.027 & 7.154$\pm$0.029 & -6.763$\pm$0.028 & 8.4 & 1.2 \\
            \hline
            \multirow{3}*{D} & XY UMa & RS CVn & 09:09:55.9355 & $+$54:29:17.724 & 0.010 & 0.010 & 14.722$\pm$0.014 & -49.781$\pm$0.012 & -182.641$\pm$0.012 & 9.4 & 1.0 \\
            & FF UMa & RS CVn & 09:33:46.5476 & $+$62:49:40.254 & 0.015 & 0.019 & 8.620$\pm$0.027 & -20.220$\pm$0.019 & -22.002$\pm$0.027 & 7.6 & 1.5 \\
            & DM UMa & RS CVn & 10:55:43.5438 & +60:28:09.721 & 0.011 & 0.011 & 5.386$\pm$0.016 & -37.912$\pm$0.014 & -7.566$\pm$0.014 & 8.9 & 1.2 \\
            \hline
            \multirow{2}*{E} & RS CVn & RS CVn & 13:10:36.9078 & $+$35:56:05.585 & 0.013 & 0.013 & 7.349$\pm$0.023 & -49.898$\pm$0.012 & 20.754$\pm$0.019 & 7.8 & 1.1 \\
            & RS UMi & RS CVn & 15:50:49.4329 & $+$72:12:40.628 & 0.011 & 0.011 & 2.119$\pm$0.011 & 3.392$\pm$0.013 & -8.534$\pm$0.014 & 9.9 & 1.0 \\
            \hline
        \end{tabular}
        \begin{tablenotes}    
            \footnotesize               
            \item[~] In the column ``Type'', ``EB'' and ``XB'' denote Eclipsing Binary and X-ray Binary, respectively. Astrometric parameters are from \textit{Gaia} DR3. $\sigma_{\alpha*}$ and $\sigma_{\delta}$ denote position errors in the RA ($\alpha\cos\delta$) and DEC directions, respectively. Column ``$G$'' gives \textit{Gaia} $G$-band magnitude. RUWE denotes \textit{Gaia} Renormalized Unit Weight Error.
        \end{tablenotes}
    \end{threeparttable}
\end{table*}

10 of the 11 selected stars are binaries except for V1859 Ori, with orbital periods ranging from 0.5 to 21 days \citep{2008MNRAS.389.1722E}.
A reliable estimation of the binary orbits requires an unaffordable amount of observing time for the ten binaries.
Fortunately, the angular scales of the binary orbits are one to two orders of magnitude smaller than the value of their parallaxes and proper motions.

\subsection{Observation setups}
\label{sect:setup}

Most radio stars are weak at radio frequencies (about millijansky level), so we requested the maximum capability of VLBA to achieve a high sensitivity: all 10 antennas, a recording rate of 4 Gbps, dual polarization, and 2-bit quantization.
The radiation of radio stars that we can detect mainly has a non-thermal nature, and their fluxes decrease as the frequency increases at centimeter wavelengths \citep{2013tra..book.....W}, so C band ($\sim$\,5\,GHz) is a good balance between flux and angular resolution.
Ionospheric propagation delay at C band is also smaller than at lower frequencies.
On the other hand, the C-band receiver provides the lowest System Equivalent Flux Density (SEFD), i.e., the highest sensitivity, among all VLBA receivers\footnote {\url{https://science.nrao.edu/facilities/vlba/docs/manuals/oss/bands-perf}}.

For weak sources, including radio stars, phase referencing \citep[PR,][]{1990AJ.....99.1663L, 1995ASPC...82..327B} using a nearby known calibrator enables precise differential astrometry.
However, atmospheric spatial structures can introduce additional propagation delays due to the angular separation between the calibrator and the target.
To achieve better atmospheric spatial structure correction compared to single-calibrator PR (hereafter referred to simply as PR) and thereby improve astrometric precision, the MultiView technique \citep{2017AJ....153..105R} is used: multiple calibrators that enclose the target are cyclically observed, allowing for a linear phase gradient estimation on the sky plane and interpolation at the target position.
The C band is suitable for the application of MultiView, e.g., \citet{2023ApJ...953...21H}.

Each target is surrounded by four calibrators, selected from the ICRF3 catalog \citep{2020A&A...644A.159C} and the Radio Fundamental Catalog \citep[RFC,][$\texttt{rfc\_2024d}$ used in this work]{2025ApJS..276...38P}\footnote{\url{https://doi.org/10.25966/dhrk-zh08}}, which are listed in Table \ref{table:calibrator}.
The calibrators are categorized into two types: primary calibrators and secondary calibrators.
The primary calibrators serve as absolute position references, meaning the phases of secondary calibrators are residual phases relative to them during the MultiView procedure, which is further introduced in Sect. \ref{sect:calib}.
As the primary calibrators are all collected from the ICRF3 catalogue, the absolute positions of the targets we measure are in the ICRF3 system.
See Appendix B of \citet{2025AJ....170....4Z} for more details about the absolute positioning.

\begin{table*}
    \centering
    \setlength{\tabcolsep}{10pt}
    \caption{Calibrators selected for each radio star}
    \label{table:calibrator}
    \begin{threeparttable}
        \begin{tabular}{llccccrr}
            \hline
            \makebox[26pt][c]{Target} & \makebox[37pt][c]{Calibrator} & RA (J2000) & DEC (J2000) & $\sigma_{\alpha*}$ & $\sigma_{\delta}$ & \makebox[35pt][c]{$S_{\mathrm{unres}}$} & \makebox[15pt][c]{Dist.} \\
            & & ($^{\mathrm{h}}$ $^{\mathrm{m}}$ $^{\mathrm{s}}$) & ($^{\mathrm{\circ}}$ $^{\mathrm{\prime}}$ $^{\mathrm{\prime\prime}}$) & (mas) & (mas) & \makebox[35pt][c]{(Jy)} & \makebox[15pt][c]{($^{\mathrm{\circ}}$)} \\
            \hline
            \multirow{4}*{FF Aqr} & J2156-0333 & 21:56:50.156791 & $-$03:33:27.94729 & 0.16 & 0.32 & 0.051 (\makebox[7pt][c]{X}) & 1.25 \\
            & J2156-0037$\star$ & 21:56:14.757927 & $-$00:37:04.59451 & 0.07 & 0.13 & 0.548 (\makebox[7pt][c]{X}) & 2.39 \\
            & J2206-0031 & 22:06:43.282610 & $-$00:31:02.49526 & 0.10 & 0.19 & 0.138 (\makebox[7pt][c]{C}) & 2.70 \\
            & J2204-0616 & 22:04:44.681846 & $-$06:16:03.38787 & 0.22 & 0.46 & 0.087 (\makebox[7pt][c]{X}) & 3.67 \\
            \hline
            \multirow{4}*{HD 8357} & J0119+0829 & 01:19:01.274274 & $+$08:29:54.70398 & 0.16 & 0.38 & 0.217 (\makebox[7pt][c]{C}) & 1.45 \\
            & J0130+0842 & 01:30:27.634426 & $+$08:42:46.17184 & 0.09 & 0.18 & 0.192 (\makebox[7pt][c]{X}) & 2.27 \\
            & J0121+0422$\star$ & 01:21:56.861693 & $+$04:22:24.73429 & 0.03 & 0.03 & 0.664 (\makebox[7pt][c]{C}) & 3.06 \\
            & J0121+1149 & 01:21:41.595041 & $+$11:49:50.41323 & 0.03 & 0.03 & 1.564 (\makebox[7pt][c]{C}) & 4.42 \\
            \hline
            \multirow{4}*{EI Eri} & J0408-0749$^{\dag}$ & 04:08:45.373699 & $-$07:49:36.07436 & 0.75 & 1.73 & 0.139 (\makebox[7pt][c]{C}) & 0.24 \\
            & J0408-0529$\star$ & 04:08:59.649940 & $-$05:29:40.53842 & 0.10 & 0.22 & 0.213 (\makebox[7pt][c]{X}) & 2.40 \\
            & J0357-0751 & 03:57:43.293242 & $-$07:51:14.56748 & 0.09 & 0.21 & 0.113 (\makebox[7pt][c]{X}) & 2.96 \\
            & J0422-0643 & 04:22:10.795426 & $-$06:43:45.33073 & 0.10 & 0.22 & 0.204 (\makebox[7pt][c]{X}) & 3.31 \\
            \hline
            \multirow{4}*{V1859 Ori} & J0519+0848 & 05:19:10.811132 & $+$08:48:56.73464 & 0.08 & 0.16 & 0.283 (\makebox[7pt][c]{X}) & 0.93 \\
            & J0517+0648$\star$ & 05:17:51.344133 & $+$06:48:03.21075 & 0.13 & 0.22 & 0.085 (\makebox[7pt][c]{X}) & 2.50 \\
            & J0532+0732 & 05:32:38.998475 & $+$07:32:43.34559 & 0.53 & 0.53 & 0.071 (\makebox[7pt][c]{C}) & 2.80 \\
            & J0521+1227$^{\dag}$ & 05:21:59.770901 & $+$12:27:05.55162 & 0.24 & 0.48 & 0.202 (\makebox[7pt][c]{C}) & 3.49 \\
            \hline
            \multirow{4}*{V1355 Ori} & J0558-0055 & 05:58:44.391585 & $-$00:55:06.92915 & 0.30 & 0.49 & 0.106 (\makebox[7pt][c]{X}) & 0.98 \\
            & J0606-0024$\star$ & 06:06:57.443582 & $-$00:24:57.46389 & 0.13 & 0.30 & 0.102 (\makebox[7pt][c]{C}) & 1.16 \\
            & J0615-0119$^{\dag}$ & 06:15:40.214886 & $-$01:19:04.66677 & 0.20 & 0.47 & 0.141 (\makebox[7pt][c]{C}) & 3.28 \\
            & J0613-0008$^{\dag}$ & 06:13:35.914480 & $-$00:08:33.74424 & 0.12 & 0.26 & 0.179 (\makebox[7pt][c]{C}) & 2.82 \\
            \hline
            \multirow{4}*{AR Mon} & J0724-0715$\star$ & 07:24:17.292627 & $-$07:15:20.35241 & 0.06 & 0.13 & 1.409 (\makebox[7pt][c]{X}) & 2.18 \\
            & J0728-0453$^{\dag}$ & 07:28:48.213248 & $-$04:53:47.77855 & 0.16 & 0.28 & 0.147 (\makebox[7pt][c]{C}) & 2.02 \\
            & J0709-0255 & 07:09:45.054609 & $-$02:55:17.49663 & 0.08 & 0.15 & 0.270 (\makebox[7pt][c]{X}) & 3.62 \\
            & J0730-0241 & 07:30:25.877613 & $-$02:41:24.90444 & 0.08 & 0.17 & 0.330 (\makebox[7pt][c]{X}) & 3.52 \\
            \hline
            \multirow{4}*{XY UMa} & J0902+5402 & 09:02:19.287489 & $+$54:02:57.25641 & 0.13 & 0.15 & 0.135 (\makebox[7pt][c]{C}) & 1.19 \\
            & J0903+5151$\star$ & 09:03:58.574560 & $+$51:51:00.66298 & 0.11 & 0.12 & 0.174 (\makebox[7pt][c]{C}) & 2.78 \\
            & J0854+5757 & 08:54:41.996416 & $+$57:57:29.93913 & 0.05 & 0.06 & 0.671 (\makebox[7pt][c]{C}) & 4.06 \\
            & J0927+5717 & 09:27:06.053451 & $+$57:17:45.34313 & 0.18 & 0.22 & 0.089 (\makebox[7pt][c]{C}) & 3.70 \\
            \hline
            \multirow{4}*{FF UMa} & J0921+6215$\star$ & 09:21:36.231076 & $+$62:15:52.18035 & 0.03 & 0.03 & 0.588 (\makebox[7pt][c]{C}) & 1.51 \\
            & J0944+6135$^{\dag}$ & 09:44:20.441650 & $+$61:35:50.19380 & 0.10 & 0.15 & 0.105 (\makebox[7pt][c]{C}) & 1.74 \\
            & J0932+6507 & 09:32:54.577582 & $+$65:07:41.29670 & 0.13 & 0.13 & 0.137 (\makebox[7pt][c]{X}) & 2.30 \\
            & J0958+6533 & 09:58:47.245113 & $+$65:33:54.81804 & 0.03 & 0.03 & 1.276 (\makebox[7pt][c]{C}) & 3.86 \\
            \hline
            \multirow{4}*{DM UMa} & J1048+6008$^{\dag}$ & 10:48:33.699074 & $+$60:08:45.64158 & 0.20 & 0.23 & 0.041 (\makebox[7pt][c]{C}) & 0.94 \\
            & J1104+6038$\star$ & 11:04:53.694641 & $+$60:38:55.31456 & 0.13 & 0.15 & 0.099 (\makebox[7pt][c]{C}) & 1.14 \\
            & J1102+5941 & 11:02:42.762830 & $+$59:41:19.58467 & 0.13 & 0.15 & 0.164 (\makebox[7pt][c]{C}) & 1.17 \\
            & J1101+6241 & 11:01:53.450783 & $+$62:41:50.60470 & 0.22 & 0.28 & 0.220 (\makebox[7pt][c]{C}) & 2.35 \\
            \hline
            \multirow{4}*{RS CVn} & J1308+3546$\star$ & 13:08:23.709135 & $+$35:46:37.16400 & 0.03 & 0.04 & 0.370 (\makebox[7pt][c]{C}) & 0.48 \\
            & J1317+3425 & 13:17:36.494184 & $+$34:25:15.93242 & 0.05 & 0.07 & 0.232 (\makebox[7pt][c]{C}) & 2.08 \\
            & J1310+3233 & 13:10:59.402733 & $+$32:33:34.44955 & 0.03 & 0.03 & 0.592 (\makebox[7pt][c]{C}) & 3.38 \\
            & J1317+3925$^{\dag}$ & 13:17:18.635659 & $+$39:25:28.14246 & 0.16 & 0.21 & 0.092 (\makebox[7pt][c]{C}) & 3.73 \\
            \hline
            \multirow{4}*{RS UMi} & J1531+7206$\star$ & 15:31:33.578594 & $+$72:06:41.22799 & 0.11 & 0.09 & 0.215 (\makebox[7pt][c]{X}) & 1.48 \\
            & J1556+7420 & 15:56:02.990517 & $+$74:20:58.14090 & 0.11 & 0.13 & 0.104 (\makebox[7pt][c]{C}) & 2.17 \\
            & J1603+6945 & 16:03:18.621508 & $+$69:45:57.44291 & 0.16 & 0.14 & 0.073 (\makebox[7pt][c]{X}) & 2.65 \\
            & J1646+7419$^{\dag}$ & 16:46:15.172466 & $+$74:19:11.06490 & 0.30 & 0.34 & 0.025 (\makebox[7pt][c]{X}) & 4.5 \\
            \hline
        \end{tabular}
        \begin{tablenotes}    
            \footnotesize
            \item[~] The calibrators with ``$\star$'' are used as the primary calibrator for sMV and the calibrator for PR.
            Most positions and uncertainties in this table are from the ICRF3 catalog, while those sources not included in ICRF3 are labeled with ``$\dag$'', and their positions and uncertainties are collected from $\texttt{rfc\_2024d}$.
            Column ``$\sigma_{\alpha*}$'' and  ``$\sigma_{\delta}$'' denote position uncertainties in the RA ($\alpha\cos\delta$) and DEC directions respectively.
            Unresolved fluxes, along with their corresponding bands in parentheses, are collected from $\texttt{rfc\_2024d}$ and listed in column ``$S_{\mathrm{unres}}$''.
            Column ``Dist.'' denotes angular distance from the target.
        \end{tablenotes}
    \end{threeparttable}
\end{table*}

\begin{figure}
    \includegraphics[width=\columnwidth]{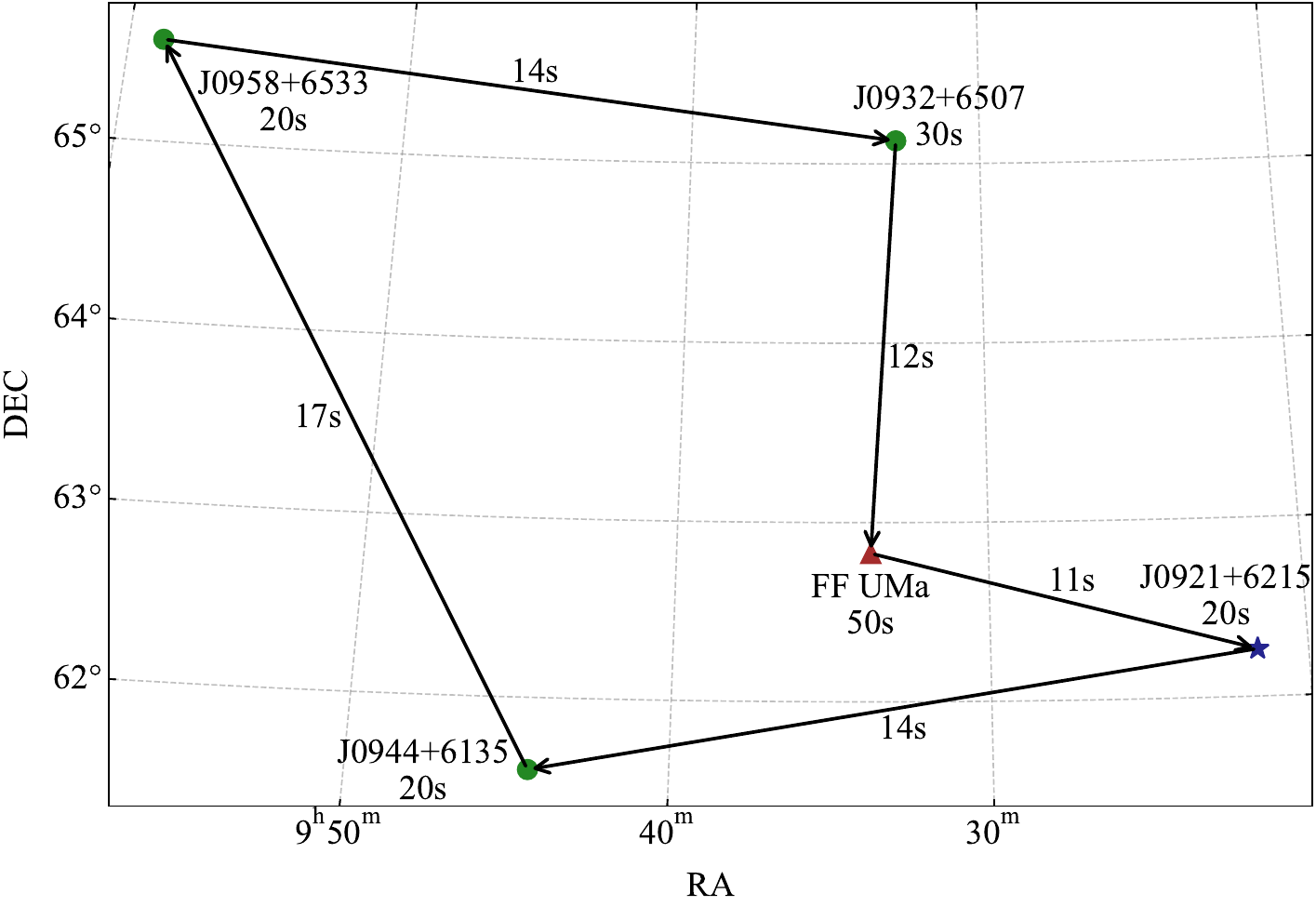}
    \caption{
        The MultiView observing cycle for FF UMa.
    }
    \label{fig:mv_cycle}
\end{figure}

The observing sequence is ``C1-C2-T-C3-C4'', where ``C$n$'' and ``T'' denote the $n$th calibrator and the target, respectively.
The observing cycle is $\sim$200 seconds, typically including 50 seconds on the target and about 20 to 30 seconds on each calibrator according to their fluxes.
An example of the observing cycle for FF UMa is shown in Fig.~\ref{fig:mv_cycle}, whose total cycle length is 209 seconds.

\subsection{Scheduling}
\label{sect:schedule}

Due to the typically low and variable fluxes of radio stars, seven epochs were scheduled for each group of targets to improve the likelihood of successfully measuring their parallaxes and proper motions.
The epochs span a period of three years, from 2021 to 2025, instead of the usual one year: BZ087 in semester 2021B, BZ103 in 2024A, and BZ107 in 2024B.
This improves the accuracy of proper motions, the most important parameter for CRF spin estimation.
The epochs are optimized to sample near the extremes of parallax sinusoids to enable more accurate parallax measurements \citep{2022PASP..134l3001R}.
The actual epochs for each session are listed in Table \ref{table:epoch}.

\begin{table*}
    \centering
    \setlength{\tabcolsep}{7pt}
    \caption{Observing epochs for each group}
    \label{table:epoch}
    \begin{threeparttable}
        \begin{tabular}{ccccccccc}
            \hline
            \multirow{2}*{Group} & \multicolumn{8}{c}{Semester (program code)} \\
             & \multicolumn{3}{c}{2021B (BZ087)} & \multicolumn{3}{c}{2024A (BZ103)} & \multicolumn{2}{c}{2024B (BZ107)} \\
            \hline
            A & 2021-11-20 & 2021-12-26 & - & 2024-06-03 & 2024-06-14 & 2024-07-03 & 2024-11-30 & 2025-01-23 \\
            B & 2021-08-19 & 2021-09-11 & 2022-01-16 & 2024-03-04 & 2024-04-03 & - & 2024-08-16 & 2024-10-07 \\
            C & 2021-09-05 & 2021-10-21 & - & 2024-03-02 & 2024-03-27 & 2024-05-04 & 2024-09-15 & 2024-10-30 \\
            D & 2021-10-09 & 2021-12-06 & - & 2024-04-14 & 2024-05-05 & 2024-06-04 & 2024-10-18 & 2024-12-07 \\
            E & 2021-08-10 & 2021-08-18 & 2021-12-13 & 2024-02-09 & 2024-03-13 & - & 2024-08-19 & 2024-09-05 \\
            \hline
        \end{tabular}
        \begin{tablenotes}
            \footnotesize
            \item The date format is YYYY-MM-DD.
        \end{tablenotes}
    \end{threeparttable}
\end{table*}

For the groups comprising two radio stars (groups A, B, C, and E), 5 hours are allocated per epoch, whereas for the group including three stars (group D), the observing time is 6.5 hours.
This ensures an on-source time of about 40 minutes for each radio star, and the corresponding theoretical thermal noise is $\sim$20\,$\mu$Jy/beam.

\section{Data reduction}
\label{sect:reduce}

\subsection{Calibration}
\label{sect:calib}

The raw data were correlated by the VLBA DiFX software correlator \citep{2007PASP..119..318D} in Socorro.
A preliminary calibration with standard PR workflow using the primary calibrator was applied to the visibility data:

\begin{enumerate}
    \item Amplitude calibration with the Astronomical Image Processing System \citep[AIPS,][]{2003ASSL..285..109G} tasks $\texttt{ACCOR}$ and $\texttt{APCAL}$, where the former automatically corrects amplitudes in cross-correlation spectra and the latter corrects based on system temperature and gain curve tables.
    \item Parallactic angle and Earth Orientation Parameters (EOP) correction with the AIPS task $\texttt{CLCOR}$.
    \item Ionospheric Faraday rotation and dispersive delay correction with the AIPS task $\texttt{TECOR}$ based on the global ionosphere maps provided by the Jet Propulsion Laboratory (JPL).
    \item Manual phase calibration with the AIPS task $\texttt{FRING}$, i.e., fringe fitting with a fringe finder to remove electronic phase differences among different intermediate frequency (IF) bands.
    \item Fringe fitting ($\texttt{FRING}$) with the primary calibrator and apply the result to the target and all secondary calibrators.
\end{enumerate}

Most of the non-spatial systematic errors were removed through the PR procedure above.
Then serial MultiView \citep[sMV,][]{2025AJ....170....4Z} was applied to remove the residual spatial-structure errors.
The sMV is an implementation of the MultiView techniques that enables automatic phase ambiguity correction.
In order to include the time-domain information and detect ``phase wrap'' robustly, the sMV operates on the time series of calibrator scans, rather than fitting independent linear phase gradients to each scan group as done in the conventional MultiView implementation.
Fringe fitting with the secondary calibrators separately yields their residual phases relative to the primary calibrator, since the fringe fitting result with the primary calibrator has already been applied to them.
Under the assumption that the residual phases change continuously, the phase plane should be highly predictable over a short time range, enabling ambiguity detection.
The phase plane is iteratively rotated to follow the changes in the residual phase time series, and then the phase correction to be applied to the target can be interpolated in the space and time domain.

We developed a pipeline based on AIPS and ParselTongue \citep{2006ASPC..351..497K}, and it is used for all data in this study.
For the data used in this work (with the observing sequence ``C1-C2-T-C3-C4''), the calibration performance of the sMV pipeline shows virtually no difference from the conventional MultiView implementation, which has been verified in \citet{2025AJ....170....4Z}.
Details of the whole procedure can also be found there.

After calibration, the targets were mapped and cleaned with the AIPS task $\texttt{IMAGR}$.
An image example is shown in Appendix \ref{app:image}.
All targets in this study are point-like sources, so their positions were fitted with the AIPS task $\texttt{JMFIT}$, assuming a single 2-D Gaussian component for each target.
All measured positions are available in Appendix \ref{app:pos}.

\subsection{Astrometric parameter estimation}
\label{sect:para_est}

For stars with $\ge$\,3 detections, position at reference epoch, parallax, and proper motion can be estimated.
Here we adopt an astrometric model of five parameters: $[\alpha, \delta, \varpi, \mu_{\alpha*}, \mu_{\delta}]$.
Binary orbit, Galactocentric acceleration, and other possible small terms are not taken into consideration.
For the parallactic motion, the DE440 \citep{2021AJ....161..105P} ephemeris provided by JPL was used to calculate precise coordinates of the Earth in the Barycentric System.
The $\texttt{emcee}$ Markov Chain Monte Carlo sampler \citep{2013PASP..125..306F} was used for parameter estimation.

Usually, errors in VLBI astrometry consist of both thermal and systematic errors \citep{2022PASP..134l3001R}.
The AIPS task $\texttt{JMFIT}$ calculates a theoretical uncertainty based on the image root mean square (RMS), leading to a significant underestimation of systematic errors.
Therefore, systematic error floors in the RA and DEC directions are iteratively added during astrometric parameter estimation to achieve a reduced chi-square $\chi^2_{\mathrm{red}}\approx 1$.
Note this can only be done for stars with $\ge$\,4 detections because the iteration procedure requires an extra degree of freedom in each direction.

As discussed in Sect. \ref{sect:target}, the angular scales of the orbits of the binaries are small.
Therefore, a single-star model is used for fitting.
And on the other hand, binaries with such short periods and small orbital scales are also fitted with single-star models in the current \textit{Gaia} astrometric solution; as discussed in \citet{2023A&A...674A...9H}, the shortest astrometric binary period ($\sim$\,30 days) successfully solved in \textit{Gaia} DR3 is larger than the longest one (AR Mon, 21 days) in our targets.

\subsection{Details of each star}
\label{sect:detail}

In general, the data quality is high, with the vast majority of calibrators being sufficiently bright and compact.
The detection rate for targets reaches $\sim 82 \%$, typically appearing as point-like sources, and the sMV gives higher SNRs than PR in most cases.
Multiple antennas were experiencing rainfall during the sessions in which sMV does not perform well.
Special cases and non-detections will be introduced below.

\subsubsection*{FF Aqr}
\label{sect:ffaqr}

FF Aqr has a slightly variable flux (the measured values for the point-like radio stars are the peak flux density $S_{\mathrm{peak}}$, same below).
There is one epoch (BZ103A2) where sMV gives a lower SNR than PR.
One of the calibrators, J2156-0333, shows a slight jet-like structure, but it does not introduce significant spurious phases into the residual phase time series.
Here, ``spurious phase'' refers to an evident phase inconsistency between this calibrator and the others on some or all baselines (same definition used below).
Self-calibrated images of J2156-0333 and other calibrators with notable structures are shown in Appendix \ref{app:structure}.

\subsubsection*{HD 8357}
\label{sect:hd8357}

The flux of HD 8357 is highly variable, varying from about 0.2\,mJy to over 20\,mJy.
There is one epoch (BZ087A2) where sMV gives a lower SNR than PR.
One of the calibrators, J0119+0829, has complex structures and shows significant spurious phases, so it is not used for sMV phase plane estimation.

\subsubsection*{EI Eri}
\label{sect:eieri}

The flux of EI Eri is highly variable, varying from about 0.5\,mJy to over 30\,mJy.

\subsubsection*{V1859 Ori}
\label{sect:v1859ori}

The flux of V1859 Ori is highly variable, varying from about 0.2\,mJy to about 5\,mJy.
One of the calibrators, J0532+0732, is not compact enough, so it does not contribute much to the long baselines (fringe fitting failure or being flagged).

\subsubsection*{V1355 Ori}
\label{sect:v1355ori}

V1355 Ori was not detected in one epoch (BZ087C1), and its flux is highly variable, varying from about 0.3\,mJy to about 5\,mJy.

\subsubsection*{AR Mon}
\label{sect:armon}

AR Mon was only detected in the first two epochs (BZ087C1, BZ087C2), which is not enough for astrometric parameter estimation.

\subsubsection*{XY UMa}
\label{sect:xyuma}

XY UMa was only detected in three epochs (BZ087D1, BZ087D2, and BZ103D2), which is enough for astrometric parameter estimation but not enough for the iterative systematic error estimation procedure (insufficient degrees of freedom, as discussed in Sect. \ref{sect:para_est}).
There is one epoch (BZ087D2) where sMV gives a lower SNR than PR.

\subsubsection*{FF UMa}
\label{sect:ffuma}

The flux of FF UMa is highly variable, varying from about 1\,mJy to over 20\,mJy.

\subsubsection*{DM UMa}
\label{sect:dmuma}

DM UMa was not detected in the last epoch (BZ107D2), and its flux is slightly variable in the first 6 epochs.
DM UMa is the only case that sMV gives lower SNRs than PR in all epochs.
Two of the secondary calibrators, J1048+6008 and J1102+5941, have jet-like structures that cause significant spurious phases.
Both of the remaining two calibrators are on the same side of DM UMa, making it impossible to interpolate phases at the position of DM UMa without J1048+6008 and J1102+5941.

\subsubsection*{RS CVn}
\label{sect:rscvn}

RS CVn was not detected in one epoch (BZ087E3), and its flux is highly variable, varying from about 0.4\,mJy to about 4\,mJy.
There are two epochs (BZ087E1, BZ107E2) where sMV gives lower SNRs than PR, and the SNRs are very close: this is due to the very small angular separation ($0.48^{\circ}$) between the primary calibrator and the target.

\subsubsection*{RS UMi}
\label{sect:rsumi}

RS UMi was not detected in two epochs (BZ087E2, BZ103E1), and its flux is highly variable, varying from about 0.3\,mJy to about 2\,mJy.
There are two epochs (BZ087E1, BZ107E2) where sMV gives lower SNRs than PR.
One of the calibrators, J1646+7419, has complex structures and shows significant spurious phases, so it is not used for sMV phase plane estimation.

\section{Results}
\label{sect:result}

\subsection{Gains of MultiView}
\label{sect:gain_lim_mv}

The MultiView techniques should outperform PR in ideal situations, that is, there are close enough calibrators enclosing the target, the calibrators are compact and have precise a priori positions, the observing cycle is within the coherence time, and most importantly, the atmosphere (and/or other spatial-structure errors) is the dominant error source.

The performance comparison between the sMV and PR can be conducted using the RMS-based $\texttt{JMFIT}$ error or the fractional flux recovery (FFR) quantity, which is defined as the ratio between the peak brightness in the astrometric image and the self-calibrated image of the same source \citep{2017AJ....153..105R}.
The radio stars are too weak for self-calibration, so here we use the ratio between the peak flux densities of PR and sMV instead.

\begin{figure}
    \includegraphics[width=\columnwidth]{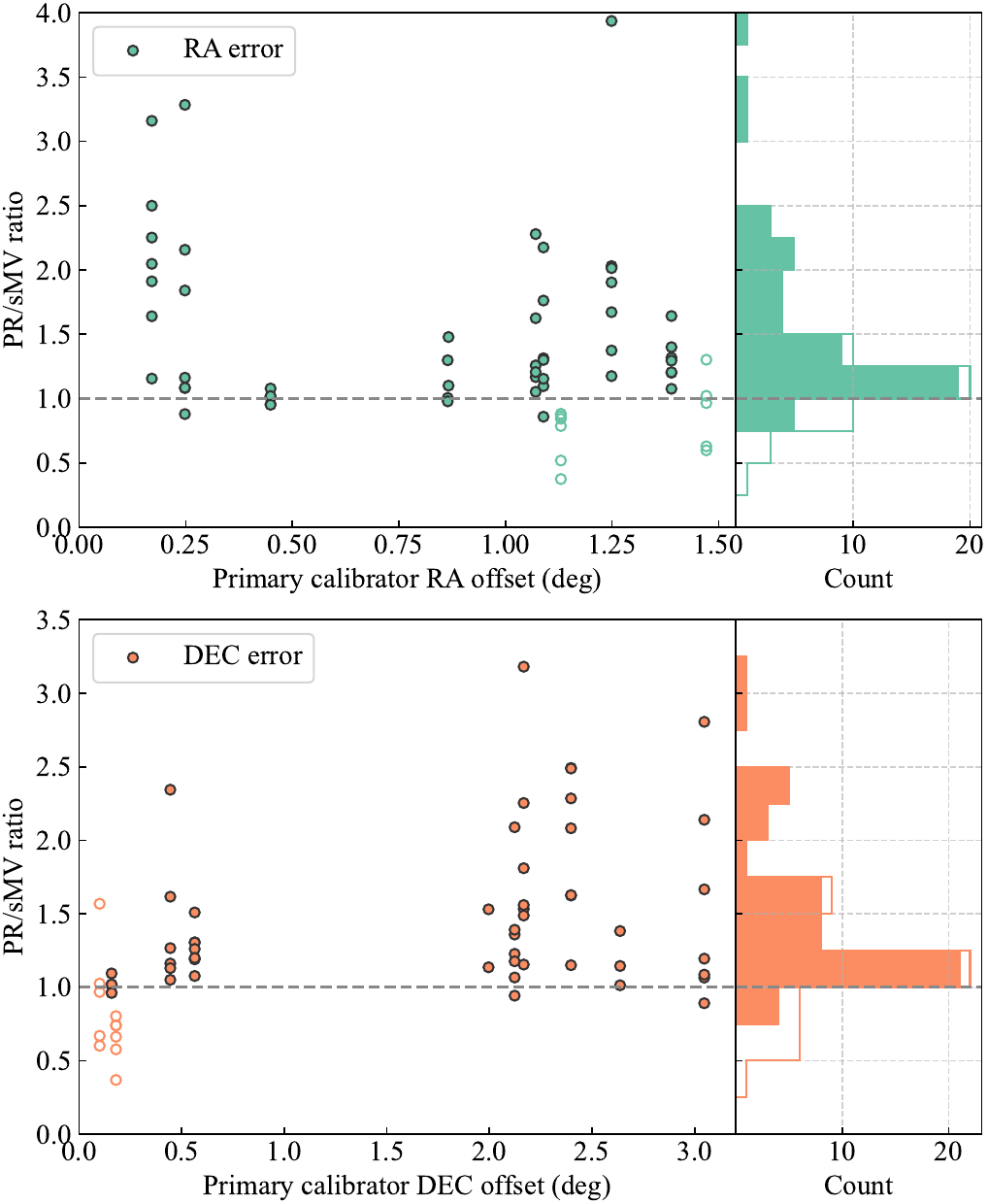}
    \caption{
        $\texttt{JMFIT}$ error ratio of PR to sMV versus the RA and DEC offsets between the target and the primary calibrator (which also serves as the PR calibrator).
        The histogram is shown on the right side.
        Upper panel: RA direction; Lower panel: DEC direction.
        RA offsets are multiplied by $\cos\delta$ of the targets.
        Hollow markers in the scatterplots and hollow areas in the histograms denote data points of DM UMa and RS UMi.
    }
    \label{fig:jmfit_ratio}
\end{figure}

Fig.~\ref{fig:jmfit_ratio} shows the $\texttt{JMFIT}$ error ratio of PR to sMV versus the RA and DEC offsets between the target and the primary calibrator of all sessions.
The sMV in most cases ($99/126$) reduces the errors (PR/sMV > 1), and it is worth noting that in cases where sMV performs poorly, DM UMa accounts for about half of them ($12/27$).
Another star with poor sMV performance is RS UMi, which, as mentioned in Sect.~\ref{sect:detail}, lost an available calibrator.

\begin{figure}
    \includegraphics[width=\columnwidth]{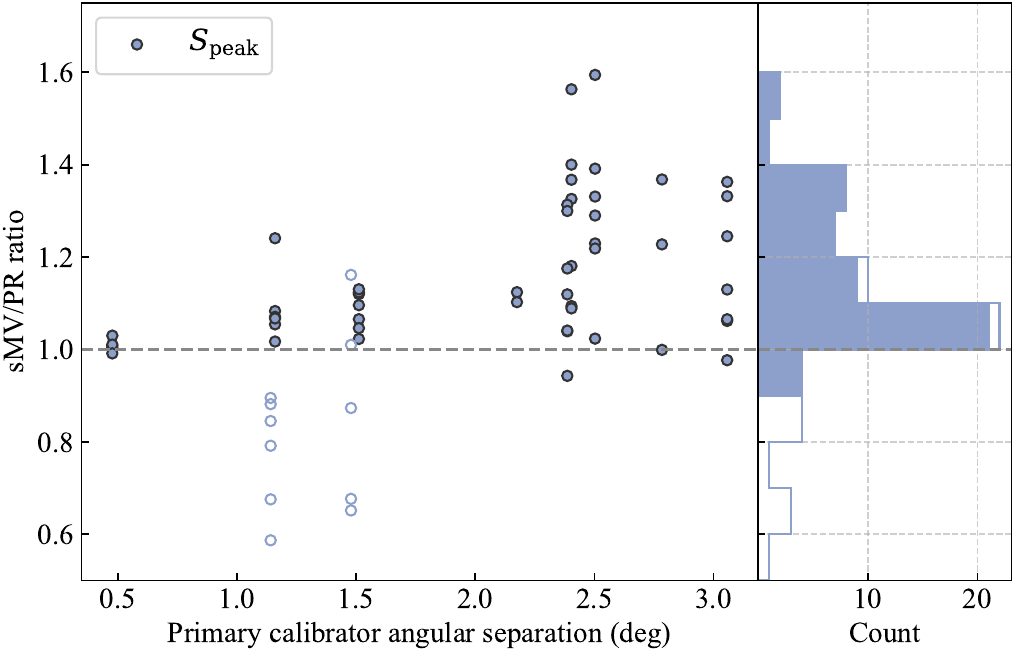}
    \caption{
        Peak flux density ratio of sMV to PR versus the angular separation between the target and the primary calibrator.
        The histogram is shown on the right side.
        Hollow markers in the scatterplot and hollow areas in the histogram denote data points of DM UMa and RS UMi.
    }
    \label{fig:error_separation}
\end{figure}

Fig.~\ref{fig:error_separation} shows the peak flux density ratio of sMV to PR versus the angular separation between the target and the primary calibrator.
Similar to the $\texttt{JMFIT}$ error, the sMV reduces flux loss in most cases (sMV/PR > 1, $50/63$), and the poor cases are mainly attributed to DM UMa and RS UMi (9/13).

The performance advantage of the sMV over PR is positively correlated with the angular separation between the target and the primary calibrator, especially in the DEC direction, as shown in Fig.~\ref{fig:jmfit_ratio} and \ref{fig:error_separation}.
This is reasonable since observations are usually conducted around its transit, and angular separations in the DEC direction correspond to differences in elevation angle.
Residual delays/phases from both the ionosphere and troposphere tend to be more pronounced in the vertical direction than in the horizontal direction, especially at low elevation angles.

\subsection{Astrometric parameters}
\label{sect:astrometry}

The estimated astrometric parameters for ten stars (excluding AR Mon) are listed in Table \ref{table:para}, and the corresponding parallax curves are presented in Appendix \ref{app:curve}.
The position uncertainties of primary calibrators are added to the position uncertainties of targets, so that the uncertainties reflect errors in absolute position rather than offsets to primary calibrators.
As discussed in Sect. \ref{sect:detail}, the uncertainties for astrometric parameters of XY UMa are unreliable.

\begin{table*}
    \centering
    \setlength{\tabcolsep}{6pt}
    \caption{Estimated astrometric parameters for ten stars}
    \label{table:para}
    \begin{threeparttable}
        \begin{tabular}{lrrrccrrr}
            \hline
            \makebox[26pt][c]{Star} & \makebox[22pt][c]{Tech.} & \makebox[60pt][c]{RA} & \makebox[60pt][c]{DEC} & $\sigma_{\alpha*}$ & $\sigma_{\delta}$ & \makebox[46pt][c]{$\varpi$} & \makebox[46pt][c]{$\mu_{\alpha*}$} & \makebox[46pt][c]{$\mu_\delta$} \\
             & & \makebox[60pt][c]{($^{\mathrm{h}}$ $^{\mathrm{m}}$ $^{\mathrm{s}}$)} & \makebox[60pt][c]{($^{\mathrm{\circ}}$ $^{\mathrm{\prime}}$ $^{\mathrm{\prime\prime}}$)} & (mas) & (mas) & \makebox[46pt][c]{(mas)} & \makebox[46pt][c]{(mas yr$^{-1}$)} & \makebox[46pt][c]{(mas yr$^{-1}$)} \\
            \hline
            \multirow{2}*{FF Aqr} & PR\makebox[6pt][c]{~} & 22:00:36.4739537 & $-$02:44:27.109490 & 0.176 & 0.338 & 4.308$\pm$0.131 & 32.740$\pm$0.083 & -11.740$\pm$0.138 \\
             & sMV\makebox[6pt][c]{$\star$} & 22:00:36.4739751 & $-$02:44:27.109979 & 0.126 & 0.212 & 4.564$\pm$0.072 & 32.774$\pm$0.043 & -11.826$\pm$0.061 \\
            \multirow{2}*{HD 8357} & PR\makebox[6pt][c]{~} & 01:22:56.9041797 & $+$07:25:14.704807 & 0.341 & 0.676 & 22.968$\pm$0.350 & 93.811$\pm$0.228 & 230.943$\pm$0.464 \\
             & sMV\makebox[6pt][c]{$\star$} & 01:22:56.9041707 & $+$07:25:14.704379 & 0.351 & 0.738 & 22.638$\pm$0.366 & 93.885$\pm$0.223 & 230.683$\pm$0.483 \\
            \multirow{2}*{EI Eri} & PR\makebox[6pt][c]{~} & 04:09:40.9526202 & $-$07:53:31.732253 & 0.298 & 0.654 & 17.727$\pm$0.229 & 36.296$\pm$0.145 & 111.943$\pm$0.316 \\
             & sMV\makebox[6pt][c]{$\star$} & 04:09:40.9526282 & $-$07:53:31.731358 & 0.229 & 0.374 & 17.894$\pm$0.146 & 36.433$\pm$0.097 & 111.925$\pm$0.109 \\
            \multirow{2}*{V1859 Ori} & PR\makebox[6pt][c]{~} & 05:22:54.7949086 & $+$08:58:04.464105 & 0.227 & 0.466 & 3.478$\pm$0.101 & 1.130$\pm$0.067 & -9.423$\pm$0.185 \\
             & sMV\makebox[6pt][c]{$\star$} & 05:22:54.7948822 & $+$08:58:04.463601 & 0.208 & 0.270 & 3.545$\pm$0.085 & 1.242$\pm$0.056 & -9.372$\pm$0.034 \\
            \multirow{2}*{V1355 Ori} & PR\makebox[6pt][c]{~} & 06:02:40.3788864 & $-$00:51:36.953470 & 0.169 & 0.369 & 7.930$\pm$0.038 & 12.484$\pm$0.033 & 11.516$\pm$0.056 \\
             & sMV\makebox[6pt][c]{$\star$} & 06:02:40.3788999 & $-$00:51:36.953609 & 0.161 & 0.374 & 7.955$\pm$0.031 & 12.514$\pm$0.028 & 11.400$\pm$0.061 \\
            \multirow{2}*{XY UMa} & PR\makebox[6pt][c]{~} & 09:09:55.8027021 & $+$54:29:13.477616 & 0.245 & 0.443 & 14.865$\pm$0.345 & -49.749$\pm$0.262 & -182.950$\pm$0.236 \\
             & sMV\makebox[6pt][c]{$\star$} & 09:09:55.8026472 & $+$54:29:13.477129 & 0.399 & 0.662 & 14.729$\pm$0.571 & -49.789$\pm$0.427 & -182.889$\pm$0.328 \\
            \multirow{2}*{FF UMa} & PR\makebox[6pt][c]{~} & 09:33:46.4791756 & $+$62:49:39.738112 & 0.107 & 0.150 & 8.432$\pm$0.079 & -20.065$\pm$0.059 & -22.223$\pm$0.086 \\
             & sMV\makebox[6pt][c]{$\star$} & 09:33:46.4791651 & $+$62:49:39.738521 & 0.128 & 0.130 & 8.470$\pm$0.098 & -20.057$\pm$0.073 & -22.157$\pm$0.072 \\
            \multirow{2}*{DM UMa} & PR\makebox[6pt][c]{$\star$} & 10:55:43.4245688 & $+$60:28:09.544893 & 0.184 & 0.213 & 5.473$\pm$0.075 & -37.848$\pm$0.050 & -7.598$\pm$0.054 \\
             & sMV\makebox[6pt][c]{~} & 10:55:43.4244712 & $+$60:28:09.543163 & 0.272 & 0.446 & 5.473$\pm$0.207 & -37.759$\pm$0.131 & -7.186$\pm$0.236 \\
            \multirow{2}*{RS CVn} & PR\makebox[6pt][c]{~} & 13:10:36.8122560 & $+$35:56:06.066910 & 0.065 & 0.221 & 7.504$\pm$0.059 & -49.952$\pm$0.026 & 20.552$\pm$0.134 \\
             & sMV\makebox[6pt][c]{$\star$} & 13:10:36.8122478 & $+$35:56:06.066804 & 0.097 & 0.181 & 7.488$\pm$0.104 & -49.973$\pm$0.051 & 20.643$\pm$0.107 \\
            \multirow{2}*{RS UMi} & PR\makebox[6pt][c]{~} & 15:50:49.4501630 & $+$72:12:40.428824 & 0.151 & 0.193 & 2.208$\pm$0.049 & 3.291$\pm$0.034 & -8.513$\pm$0.083 \\
             & sMV\makebox[6pt][c]{$\star$} & 15:50:49.4500275 & $+$72:12:40.428941 & 0.127 & 0.327 & 2.187$\pm$0.014 & 3.340$\pm$0.017 & -8.684$\pm$0.180 \\
            \hline
        \end{tabular}
        \begin{tablenotes}
            \footnotesize
            \item The reference epoch is J2023.25. Column ``Tech.'' denotes calibration technique, PR or sMV. The calibration technique with ``$\star$'' is adopted as the final astrometric result for each star. Correlation coefficient matrices of the adopted results are listed in Appendix \ref{app:corr}.
        \end{tablenotes}
    \end{threeparttable}
\end{table*}

The performance advantage of sMV over PR is related to the angular separation between the target and the primary calibrator.
For stars with a primary calibrator located within $\sim 1.5^{\circ}$ (V1355 Ori, FF UMa, RS CVn, RS UMi), the sMV and PR uncertainties are at the same level.
While for the other stars, the sMV shows a significant advantage over PR, with two exceptions, HD 8357 and DM UMa.

HD 8357 has a relatively large angular scale of its binary orbit.
The semimajor axis projected on the radial direction of its active component $a\sin i=4.9\times10^6$\,km, where $a$ is the semimajor axis and $i$ is the inclination \citep{1996AJ....112..269F}.
Calculating with its \textit{Gaia} parallax $\varpi=21.981$\,mas, the corresponding maximum angular orbital offset would be no less than 0.72\,mas, which is non-negligible.
Therefore, unless a binary model is adopted, it is difficult to achieve further improvements at the current level of accuracy.

As mentioned in Sect. \ref{sect:detail}, only two available calibrators are available for DM UMa, and they are both located on the same side of DM UMa in the sky, which leads to a MultiView failure.
Therefore, we adopt the PR result as the final astrometric result for DM UMa, whereas the sMV results are adopted for all other stars.

\begin{figure}
    \includegraphics[width=0.99\columnwidth]{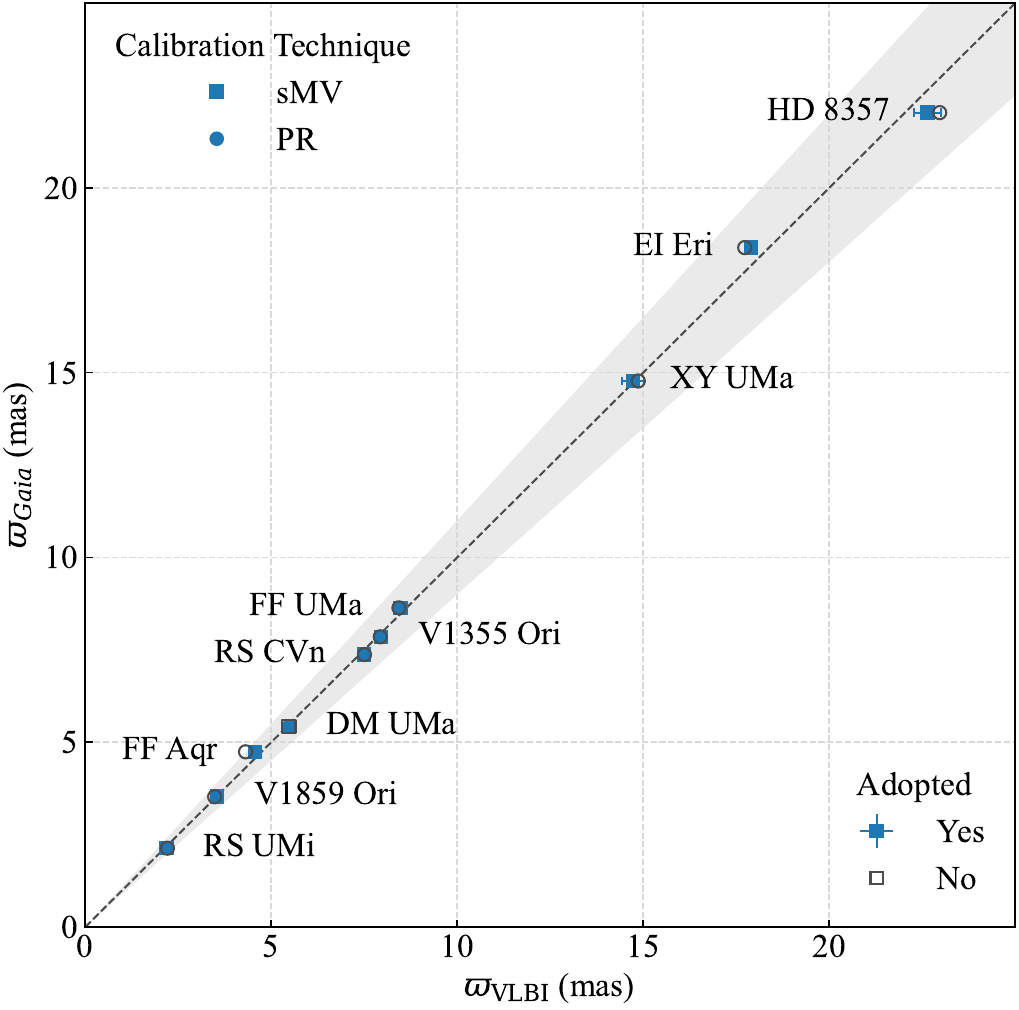}
    \caption{
        Comparison between the VLBI and \textit{Gaia} parallaxes.
        Square and circle markers denote sMV and PR, respectively.
        Filled markers denote adopted results, and error bars are plotted for them.
        The reference dashed line has a slope of 1 and an intercept of 0.
        The filled area indicates a deviation range within $\pm10\%$.
    }
    \label{fig:plx_comp}
\end{figure}

Overall, a high precision is achieved in most adopted astrometric results.
Take parallaxes as an example, the mean parallax uncertainty is $0.156$\,mas, while the median is $0.091$\,mas.
Parallax is a parameter that is not strongly affected by the systematic bright \textit{Gaia}-CRF rotation, so a direct comparison between VLBI and \textit{Gaia} is possible.
Fig. \ref{fig:plx_comp} shows the comparison between the results and \textit{Gaia} DR3 parallaxes, for which the zero-point issue has been corrected using the recipe and coefficients provided by \citet{2021A&A...649A...4L} and \citet{2024AA...691A..81D}, respectively.

Define the normalized residual between VLBI and \textit{Gaia} parallaxes as
\begin{equation}
    Z_{\varpi}=\frac{\varpi_{\mathrm{VLBI}}-\varpi_{Gaia}}{\sqrt{\sigma_{\varpi\mathrm{VLBI}}^2 + \sigma_{\varpi Gaia}^2}} \ ,
\end{equation}
and its absolute value $|Z_{\varpi}|$ can be calculated for each star, as shown in Table \ref{table:error}, together with other error statistics.
We compared the normalized residuals with and without \textit{Gaia} parallax zero-point correction, and the correction only reduces the mean absolute value of normalized residual by $\sim 0.2$ ($\langle|Z_{\varpi}|\rangle = 1.847$, $\langle|Z^{\mathrm{corr}}_{\varpi}|\rangle = 1.639$).
Assuming that the VLBI and \textit{Gaia} measurements are independent, that their uncertainties are reasonable estimates, and that no additional systematic errors exist, $Z_{\varpi}$ should follow a standard Gaussian distribution $\mathcal{N}(0, 1)$.
Therefore, the probability that $|Z_{\varpi}|<2$ is $95\%$.
However, only six of the ten stars have $|Z_{\varpi}|<2$, suggesting that one or both of VLBI and \textit{Gaia} may have underestimated their uncertainties, or there exist systematic errors.

\begin{table}
    \centering
    \setlength{\tabcolsep}{4pt}
    \caption{Error statistics of the adopted parallaxes and proper motions}
    \label{table:error}
    \begin{threeparttable}
        \begin{tabular}{lrrrrr}
            \hline
            \makebox[30pt][c]{Star} & $\sigma_{\varpi\mathrm{VLBI}}$ & \makebox[35pt][c]{$\sigma_{\mu_{\alpha*}\mathrm{VLBI}}$} & \makebox[35pt][c]{$\sigma_{\mu_{\delta}\mathrm{VLBI}}$} & \makebox[18pt][c]{$|Z_{\varpi}|$} & \makebox[18pt][c]{$|Z^{\mathrm{corr}}_{\varpi}|$} \\
            & \makebox[28pt][c]{(mas)} & (mas yr$^{-1}$) & (mas yr$^{-1}$) & & \\
            \hline
            FF Aqr & 0.072 & 0.043 & 0.061 & 2.144 & 2.212 \\
            HD 8357 & 0.366 & 0.223 & 0.483 & 1.782 & 1.638 \\
            EI Eri & 0.146 & 0.097 & 0.109 & 3.213 & 3.229 \\
            V1859 Ori & 0.085 & 0.056 & 0.034 & 0.803 & 0.240 \\
            V1355 Ori & 0.031 & 0.028 & 0.061 & 2.910 & 2.646 \\
            XY UMa & 0.571 & 0.427 & 0.328 & 0.011 & 0.078 \\
            FF UMa & 0.098 & 0.073 & 0.072 & 1.471 & 1.619 \\
            DM UMa & 0.075 & 0.050 & 0.054 & 1.134 & 0.664 \\
            RS CVn & 0.104 & 0.051 & 0.107 & 1.300 & 1.099 \\
            RS UMi & 0.014 & 0.017 & 0.180 & 3.696 & 2.961 \\
            \hline
            Mean & 0.156 & 0.106 & 0.149 & 1.847 & 1.639 \\
            Median & 0.091 & 0.053 & 0.089 & 1.627 & 1.628 \\
            \hline
        \end{tabular}
        \begin{tablenotes}
            \footnotesize
            \item Columns 2\,-\,4 list the uncertainties of the adopted parallaxes and proper motions. Column $|Z_{\varpi}|$ denotes the absolute value of the normalized residual between VLBI and \textit{Gaia} DR3 parallaxes, while $|Z^{\mathrm{corr}}_{\varpi}|$ is the normalized residual for \textit{Gaia} DR3 parallaxes after zero-point correction.
        \end{tablenotes}
    \end{threeparttable}
\end{table}

Due to the geometry of the VLBA network and the declinations of the targets, the restored synthesis beams for most targets are ellipses with approximately north-south-oriented long axes.
As a result, the angular resolution in the DEC direction is not as good as that in the RA direction, and the uncertainties of the positions and proper motions in the DEC direction are worse.

\section{Summary}
\label{sect:summary}

We observed 11 radio stars with the VLBA at C band.
With 7 epochs spanning over 3 years, all 11 stars were detected at least twice.
The data were calibrated using our pipeline with two techniques: single-calibrator phase referencing and serial MultiView.
We successfully derived parallaxes and proper motions for 10 of them, achieving median uncertainties better than 0.1\,mas and 0.1\,mas yr$^{-1}$, respectively.
The high-quality data produced in this study are an important addition to the available radio star sample for the bright \textit{Gaia}-CRF validation, and the frame link results are presented in \citet{2025A&A...699A.345Z}.

\section*{Acknowledgements}

This work is supported by the Strategic Priority Research Program of the Chinese Academy of Sciences, Grant No. XDA0350205, the National Natural Science Foundation of China (NSFC) under grant No. U2031212, and the National Key R\&D Program of China (No. 2024YFA1611501).
J. Zhang is supported by the Postdoctoral Programme for Research Institutes in Finland funded by the Finnish Government.
W. Chen is supported by Yunnan Fundamental Research Projects (grant No. 202401AT070144) and Yunnan Foreign Talent Introduction Program (grant No. 202505AO120021).

The Python pipeline used for calibration can be found at the Astrophysics Source Code Library \citep{2025ascl.soft08016Z}, and a ``frozen'' version used in this paper is available at Zenodo \citep{jingdong_zhang_2025_15030432}.
The Python program used for astrometric parameter estimation can be found at \url{https://github.com/FrdCHK/mcmc_ppm_estimate}, and a ``frozen'' version used in this paper is available at Zenodo \citep{jingdong_zhang_2025_15031700}.

The RFC can be accessed through \url{https://doi.org/10.25966/dhrk-zh08}.
This work has also made use of the SIMBAD database, operated at CDS, Strasbourg, France \citep[\url{https://simbad.u-strasbg.fr/simbad/}]{simbad}.

The European VLBI Network is a joint facility of independent European, African, Asian, and North American radio astronomy institutes. Scientific results from data presented in this publication are derived from the following EVN project codes: EZ029, EX009.

This work has made use of data from the European Space Agency (ESA) mission
\textit{Gaia} (\url{https://www.cosmos.esa.int/gaia}), processed by the \textit{Gaia}
Data Processing and Analysis Consortium (DPAC,
\url{https://www.cosmos.esa.int/web/gaia/dpac/consortium}). Funding for the DPAC
has been provided by national institutions, in particular the institutions
participating in the \textit{Gaia} Multilateral Agreement.
The \textit{Gaia} services (\url{https://gaia.ari.uni-heidelberg.de/index.html}) provided by the Astronomisches Rechen-Institut (ARI) of the University of Heidelberg are used in \textit{Gaia} data retrieval.
The Python package for \textit{Gaia} DR3 parallax zero-point correction developed by P. Ramos can be found at \url{https://gitlab.com/icc-ub/public/gaiadr3_zeropoint}, and the coefficients used in this paper can be found at \url{https://github.com/yedings/Parallax-bias-correction-in-the-Galactic-plane}.

This research has made use of the Astrophysics Data System, funded by NASA under Cooperative Agreement 80NSSC21M0056.

Software and Python packages used in this work (in alphabetical order):
AIPS \citep{2003ASSL..285..109G}, Astropy \citep{astropy2013,astropy2018,astropy2022}, Difmap \citep{1997ASPC..125...77S}, emcee \citep{2013PASP..125..306F}, Matplotlib \citep{2007CSE.....9...90H}, Numpy \citep{harris2020array}, Pandas \citep{2022zndo...3509134T}, ParselTongue \citep{2006ASPC..351..497K}, Scipy \citep{2020NatMe..17..261V}, and TOPCAT \citep{2005ASPC..347...29T}.

\section*{Data Availability}

The EVN data used in this study can be downloaded from \url{https://archive.jive.nl/scripts/portal.php}.
The VLBA data used in this study can be downloaded from the data archive of the National Radio Astronomy Observatory (NRAO; \url{https://data.nrao.edu}).
VLBI images and parallax curves are available at Zenodo \citep{jingdong_zhang_2025_17348486}: \url{https://doi.org/10.5281/zenodo.17348486}.



\bibliographystyle{mnras}
\bibliography{ref} 




\appendix



\section{VLBI images of radio stars}
\label{app:image}

\begin{figure*}
    \includegraphics[width=0.8\textwidth]{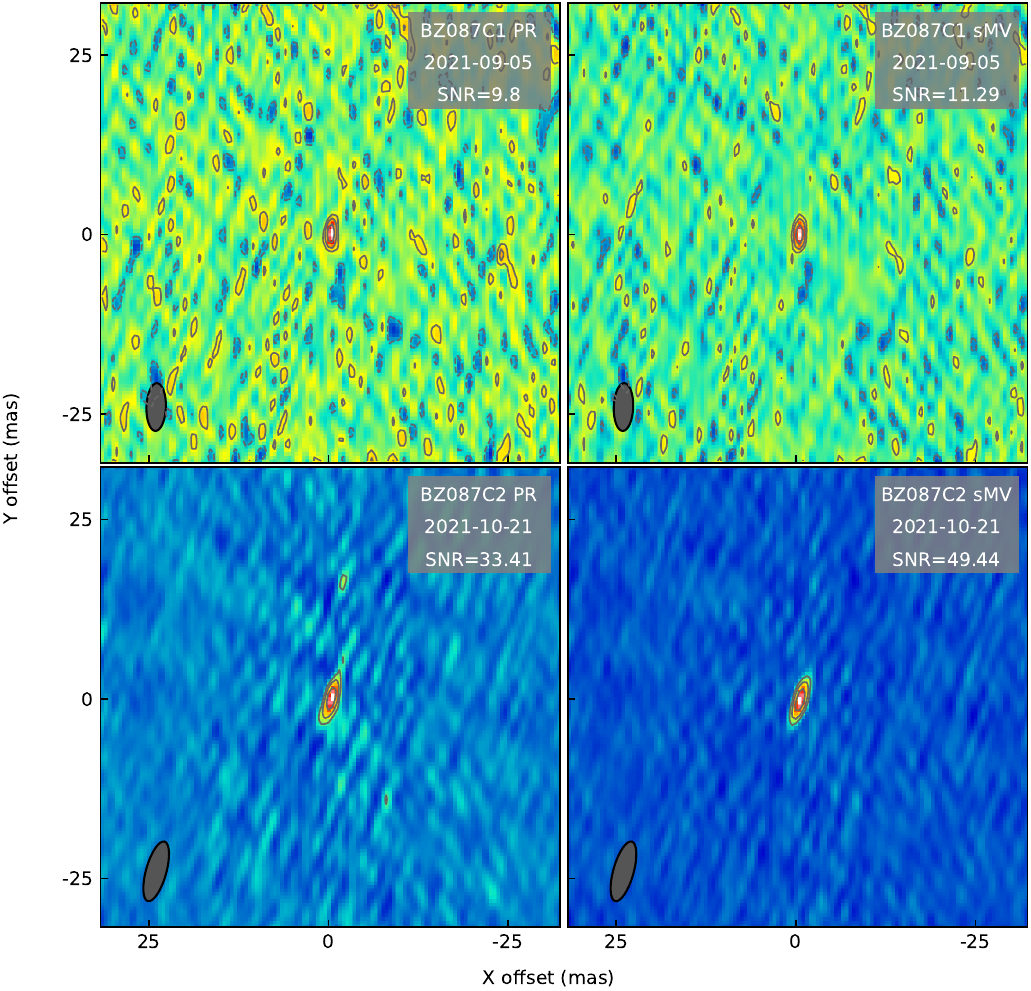}
    \caption{
        VLBI images of AR Mon.
        Left: calibrated with PR; Right: calibrated with sMV.
        Restored beams are shown at the bottom left corner of each panel.
        Contours \%: -16 16 32 64.
        All images are cleaned with the AIPS task $\texttt{IMAGR}$.
    }
    \label{fig:armon}
\end{figure*}

Here we put an example (AR Mon) of VLBI images in Fig.~\ref{fig:armon}.
Images of the remaining stars are available at Zenodo \citep{jingdong_zhang_2025_17348486}.

\section{Measured positions of radio stars}
\label{app:pos}

The successfully measured positions of all epochs of all stars are listed in Table \ref{table:pos_epoch}.
\clearpage
\onecolumn
\begin{ThreePartTable}
    \setlength{\tabcolsep}{5pt}
    \begin{longtable}{lccrrrcccc}
        \caption{Measured positions of all epochs of all stars} \label{table:pos_epoch} \\
        \hline
        \makebox[26pt][c]{Star} & Session & Epoch & Tech. & \makebox[45pt][c]{$S_{\mathrm{peak}}$} & SNR & \makebox[45pt][c]{RA / $\Delta\alpha_{*}$} & \makebox[43pt][c]{DEC / $\Delta\delta$} & $\sigma_{\alpha*}$ & $\sigma_{\delta}$ \\
         & & & & \makebox[40pt][c]{(mJy Beam$^{-1}$)} & & \makebox[45pt][c]{($^{\mathrm{h}}$ $^{\mathrm{m}}$ $^{\mathrm{s}}$) / (mas)} & \makebox[43pt][c]{($^{\mathrm{\circ}}$ $^{\mathrm{\prime}}$ $^{\mathrm{\prime\prime}}$) / (mas)} & (mas) & (mas) \\
        \hline
        \endfirsthead
        \caption{continued}\\
        \hline
        \makebox[26pt][c]{Star} & Session & Epoch & Tech. & \makebox[45pt][c]{$S_{\mathrm{peak}}$} & SNR & \makebox[45pt][c]{RA / $\Delta\alpha_{*}$} & \makebox[43pt][c]{DEC / $\Delta\delta$} & $\sigma_{\alpha*}$ & $\sigma_{\delta}$ \\
         & & & & \makebox[40pt][c]{(mJy Beam$^{-1}$)} & & \makebox[45pt][c]{($^{\mathrm{h}}$ $^{\mathrm{m}}$ $^{\mathrm{s}}$) / (mas)} & \makebox[43pt][c]{($^{\mathrm{\circ}}$ $^{\mathrm{\prime}}$ $^{\mathrm{\prime\prime}}$) / (mas)} & (mas) & (mas) \\
        \hline
        \endhead
        \hline
        \endfoot
        \hline
        \endlastfoot
        \multirow{14}*{FF Aqr} & \multirow{2}*{BZ087A1} & \multirow{2}*{2021.887909} & PR & 1.803$\pm$0.058 & 31.2 & 22:00:36.4706982 & $-$02:44:27.095425 & 0.023 & 0.051 \\
         & & & sMV & 2.368$\pm$0.042 & 56.6 & 22:00:36.4707151 & $-$02:44:27.095474 & 0.011 & 0.024 \\
         & \multirow{2}*{BZ087A2} & \multirow{2}*{2021.986277} & PR & 1.958$\pm$0.094 & 20.8 & \makebox[20pt][r]{4.486} & \makebox[20pt][r]{-0.396} & 0.051 & 0.106 \\
         & & & sMV & 2.035$\pm$0.074 & 27.3 & \makebox[20pt][r]{4.317} & \makebox[20pt][r]{-1.297} & 0.039 & 0.078 \\
         & \multirow{2}*{BZ103A1} & \multirow{2}*{2024.421763} & PR & 0.657$\pm$0.033 & 20.1 & \makebox[20pt][r]{91.181} & \makebox[20pt][r]{-26.144} & 0.041 & 0.086 \\
         & & & sMV & 0.683$\pm$0.032 & 21.6 & \makebox[20pt][r]{91.797} & \makebox[20pt][r]{-26.475} & 0.038 & 0.080 \\
         & \multirow{2}*{BZ103A2} & \multirow{2}*{2024.451798} & PR & 0.527$\pm$0.041 & 12.9 & \makebox[20pt][r]{91.773} & \makebox[20pt][r]{-25.916} & 0.113 & 0.153 \\
         & & & sMV & 0.497$\pm$0.041 & 12.1 & \makebox[20pt][r]{92.579} & \makebox[20pt][r]{-26.688} & 0.131 & 0.162 \\
         & \multirow{2}*{BZ103A3} & \multirow{2}*{2024.503675} & PR & 0.586$\pm$0.040 & 14.7 & \makebox[20pt][r]{92.568} & \makebox[20pt][r]{-27.052} & 0.065 & 0.119 \\
         & & & sMV & 0.656$\pm$0.037 & 17.6 & \makebox[20pt][r]{93.127} & \makebox[20pt][r]{-27.667} & 0.050 & 0.097 \\
         & \multirow{2}*{BZ107A1} & \multirow{2}*{2024.915961} & PR & 0.711$\pm$0.046 & 15.3 & \makebox[20pt][r]{99.308} & \makebox[20pt][r]{-35.608} & 0.068 & 0.128 \\
         & & & sMV & 0.925$\pm$0.043 & 21.3 & \makebox[20pt][r]{99.622} & \makebox[20pt][r]{-36.072} & 0.038 & 0.092 \\
         & \multirow{2}*{BZ107A2} & \multirow{2}*{2025.063430} & PR & 0.704$\pm$0.037 & 18.8 & \makebox[20pt][r]{106.841} & \makebox[20pt][r]{-36.825} & 0.032 & 0.081 \\
         & & & sMV & 0.827$\pm$0.036 & 22.9 & \makebox[20pt][r]{106.867} & \makebox[20pt][r]{-37.492} & 0.027 & 0.069 \\
         \hline
         \multirow{14}*{HD 8357} & \multirow{2}*{BZ087A1} & \multirow{2}*{2021.888136} & PR & 0.174$\pm$0.028 & 6.2 & 01:22:56.8947396 & $+$07:25:14.386772 & 0.158 & 0.279 \\
         & & & sMV & 0.185$\pm$0.029 & 6.5 & 01:22:56.8947436 & $+$07:25:14.386569 & 0.146 & 0.262 \\
         & \multirow{2}*{BZ087A2} & \multirow{2}*{2021.986504} & PR & 0.269$\pm$0.047 & 5.7 & \makebox[20pt][r]{1.711} & \makebox[20pt][r]{16.232} & 0.148 & 0.307 \\
         & & & sMV & 0.263$\pm$0.046 & 5.6 & \makebox[20pt][r]{1.636} & \makebox[20pt][r]{16.677} & 0.169 & 0.344 \\
         & \multirow{2}*{BZ103A1} & \multirow{2}*{2024.421990} & PR & 0.159$\pm$0.026 & 6.2 & \makebox[20pt][r]{268.099} & \makebox[20pt][r]{595.789} & 0.336 & 0.710 \\
         & & & sMV & 0.180$\pm$0.025 & 7.1 & \makebox[20pt][r]{267.721} & \makebox[20pt][r]{594.452} & 0.289 & 0.595 \\
         & \multirow{2}*{BZ103A2} & \multirow{2}*{2024.452024} & PR & 1.267$\pm$0.104 & 12.2 & \makebox[20pt][r]{271.346} & \makebox[20pt][r]{602.984} & 0.193 & 0.391 \\
         & & & sMV & 1.688$\pm$0.072 & 23.4 & \makebox[20pt][r]{270.963} & \makebox[20pt][r]{601.279} & 0.090 & 0.183 \\
         & \multirow{2}*{BZ103A3} & \multirow{2}*{2024.503901} & PR & 0.220$\pm$0.031 & 7.1 & \makebox[20pt][r]{279.519} & \makebox[20pt][r]{615.613} & 0.250 & 0.635 \\
         & & & sMV & 0.234$\pm$0.031 & 7.5 & \makebox[20pt][r]{279.482} & \makebox[20pt][r]{615.045} & 0.231 & 0.585 \\
         & \multirow{2}*{BZ107A1} & \multirow{2}*{2024.916188} & PR & 20.036$\pm$0.822 & 24.4 & \makebox[20pt][r]{280.829} & \makebox[20pt][r]{695.850} & 0.039 & 0.075 \\
         & & & sMV & 27.305$\pm$0.464 & 58.8 & \makebox[20pt][r]{281.099} & \makebox[20pt][r]{695.203} & 0.012 & 0.027 \\
         & \multirow{2}*{BZ107A2} & \multirow{2}*{2025.063657} & PR & 1.858$\pm$0.066 & 28.2 & \makebox[20pt][r]{290.523} & \makebox[20pt][r]{730.221} & 0.034 & 0.054 \\
         & & & sMV & 2.313$\pm$0.053 & 43.9 & \makebox[20pt][r]{290.752} & \makebox[20pt][r]{730.125} & 0.018 & 0.032 \\
         \hline
         \multirow{14}*{EI Eri} & \multirow{2}*{BZ087B1} & \multirow{2}*{2021.631932} & PR & 2.292$\pm$0.096 & 23.9 & 04:09:40.9497982 & $-$07:53:31.909620 & 0.062 & 0.111 \\
         & & & sMV & 2.706$\pm$0.069 & 38.9 & 04:09:40.9498351 & $-$07:53:31.908786 & 0.032 & 0.068 \\
         & \multirow{2}*{BZ087B2} & \multirow{2}*{2021.694736} & PR & 6.912$\pm$0.173 & 39.9 & \makebox[20pt][r]{3.006} & \makebox[20pt][r]{3.141} & 0.023 & 0.051 \\
         & & & sMV & 7.561$\pm$0.121 & 62.3 & \makebox[20pt][r]{2.683} & \makebox[20pt][r]{5.045} & 0.014 & 0.031 \\
         & \multirow{2}*{BZ087B3} & \multirow{2}*{2022.041508} & PR & 13.503$\pm$0.525 & 25.7 & \makebox[20pt][r]{-15.763} & \makebox[20pt][r]{35.493} & 0.032 & 0.072 \\
         & & & sMV & 17.904$\pm$0.324 & 55.2 & \makebox[20pt][r]{-16.006} & \makebox[20pt][r]{35.515} & 0.013 & 0.029 \\
         & \multirow{2}*{BZ103B1} & \multirow{2}*{2024.174030} & PR & 1.372$\pm$0.094 & 14.6 & \makebox[20pt][r]{58.589} & \makebox[20pt][r]{280.168} & 0.091 & 0.244 \\
         & & & sMV & 1.876$\pm$0.068 & 27.8 & \makebox[20pt][r]{58.760} & \makebox[20pt][r]{280.690} & 0.040 & 0.107 \\
         & \multirow{2}*{BZ103B2} & \multirow{2}*{2024.255828} & PR & 0.529$\pm$0.035 & 15.0 & \makebox[20pt][r]{65.091} & \makebox[20pt][r]{294.059} & 0.070 & 0.167 \\
         & & & sMV & 0.576$\pm$0.033 & 17.4 & \makebox[20pt][r]{65.253} & \makebox[20pt][r]{294.592} & 0.061 & 0.145 \\
         & \multirow{2}*{BZ107B1} & \multirow{2}*{2024.624481} & PR & 3.324$\pm$0.228 & 14.6 & \makebox[20pt][r]{109.342} & \makebox[20pt][r]{336.315} & 0.077 & 0.181 \\
         & & & sMV & 4.655$\pm$0.151 & 30.9 & \makebox[20pt][r]{109.647} & \makebox[20pt][r]{336.713} & 0.037 & 0.087 \\
         & \multirow{2}*{BZ107B2} & \multirow{2}*{2024.766461} & PR & 20.223$\pm$1.180 & 17.2 & \makebox[20pt][r]{109.758} & \makebox[20pt][r]{342.171} & 0.076 & 0.128 \\
         & & & sMV & 31.612$\pm$0.883 & 35.8 & \makebox[20pt][r]{110.307} & \makebox[20pt][r]{344.177} & 0.024 & 0.052 \\
         \hline
         \multirow{14}*{V1859 Ori} & \multirow{2}*{BZ087B1} & \multirow{2}*{2021.632072} & PR & 3.806$\pm$0.149 & 25.6 & 05:22:54.7950173 & $+$08:58:04.479906 & 0.052 & 0.076 \\
         & & & sMV & 4.679$\pm$0.055 & 85.5 & 05:22:54.7949568 & $+$08:58:04.479273 & 0.013 & 0.024 \\
         & \multirow{2}*{BZ087B2} & \multirow{2}*{2021.694879} & PR & 1.265$\pm$0.033 & 38.2 & \makebox[20pt][r]{-0.200} & \makebox[20pt][r]{-1.039} & 0.021 & 0.045 \\
         & & & sMV & 1.295$\pm$0.029 & 44.0 & \makebox[20pt][r]{-0.444} & \makebox[20pt][r]{-1.419} & 0.018 & 0.039 \\
         & \multirow{2}*{BZ087B3} & \multirow{2}*{2022.041651} & PR & 0.139$\pm$0.036 & 3.8 & \makebox[20pt][r]{-4.729} & \makebox[20pt][r]{-4.677} & 0.405 & 0.550 \\
         & & & sMV & 0.221$\pm$0.035 & 6.3 & \makebox[20pt][r]{-5.066} & \makebox[20pt][r]{-5.606} & 0.200 & 0.359 \\
         & \multirow{2}*{BZ103B1} & \multirow{2}*{2024.174174} & PR & 1.188$\pm$0.066 & 18.0 & \makebox[20pt][r]{-4.064} & \makebox[20pt][r]{-25.400} & 0.056 & 0.106 \\
         & & & sMV & 1.581$\pm$0.046 & 34.4 & \makebox[20pt][r]{-4.522} & \makebox[20pt][r]{-25.226} & 0.030 & 0.058 \\
         & \multirow{2}*{BZ103B2} & \multirow{2}*{2024.255971} & PR & 0.555$\pm$0.035 & 15.7 & \makebox[20pt][r]{-3.663} & \makebox[20pt][r]{-25.689} & 0.058 & 0.103 \\
         & & & sMV & 0.676$\pm$0.031 & 21.6 & \makebox[20pt][r]{-4.076} & \makebox[20pt][r]{-25.745} & 0.035 & 0.069 \\
         & \multirow{2}*{BZ107B1} & \multirow{2}*{2024.624624} & PR & 1.260$\pm$0.070 & 18.0 & \makebox[20pt][r]{2.951} & \makebox[20pt][r]{-27.188} & 0.055 & 0.162 \\
         & & & sMV & 1.753$\pm$0.052 & 33.7 & \makebox[20pt][r]{2.834} & \makebox[20pt][r]{-28.645} & 0.027 & 0.072 \\
         & \multirow{2}*{BZ107B2} & \multirow{2}*{2024.766604} & PR & 0.347$\pm$0.033 & 10.4 & \makebox[20pt][r]{3.394} & \makebox[20pt][r]{-30.231} & 0.071 & 0.197 \\
         & & & sMV & 0.447$\pm$0.033 & 13.6 & \makebox[20pt][r]{3.294} & \makebox[20pt][r]{-30.444} & 0.051 & 0.126 \\
         \pagebreak
         \multirow{12}*{V1355 Ori} & \multirow{2}*{BZ087C2} & \multirow{2}*{2021.804178} & PR & 1.224$\pm$0.034 & 35.6 & 06:02:40.3781456 & $-$00:51:36.971648 & 0.022 & 0.046 \\
         & & & sMV & 1.245$\pm$0.033 & 37.5 & 06:02:40.3781582 & $-$00:51:36.971629 & 0.021 & 0.044 \\
         & \multirow{2}*{BZ103C1} & \multirow{2}*{2024.166067} & PR & 4.894$\pm$0.136 & 35.9 & \makebox[20pt][r]{15.153} & \makebox[20pt][r]{27.843} & 0.027 & 0.063 \\
         & & & sMV & 5.301$\pm$0.084 & 62.9 & \makebox[20pt][r]{15.192} & \makebox[20pt][r]{27.492} & 0.016 & 0.039 \\
         & \multirow{2}*{BZ103C2} & \multirow{2}*{2024.236962} & PR & 1.190$\pm$0.041 & 29.2 & \makebox[20pt][r]{15.405} & \makebox[20pt][r]{29.910} & 0.032 & 0.078 \\
         & & & sMV & 1.254$\pm$0.033 & 38.2 & \makebox[20pt][r]{15.730} & \makebox[20pt][r]{29.822} & 0.025 & 0.061 \\
         & \multirow{2}*{BZ103C3} & \multirow{2}*{2024.340697} & PR & 0.896$\pm$0.040 & 22.2 & \makebox[20pt][r]{18.907} & \makebox[20pt][r]{32.939} & 0.038 & 0.076 \\
         & & & sMV & 0.959$\pm$0.037 & 25.6 & \makebox[20pt][r]{19.202} & \makebox[20pt][r]{32.624} & 0.033 & 0.065 \\
         & \multirow{2}*{BZ107C1} & \multirow{2}*{2024.706567} & PR & 0.399$\pm$0.030 & 13.4 & \makebox[20pt][r]{37.170} & \makebox[20pt][r]{35.209} & 0.047 & 0.106 \\
         & & & sMV & 0.426$\pm$0.029 & 14.9 & \makebox[20pt][r]{37.482} & \makebox[20pt][r]{34.889} & 0.039 & 0.093 \\
         & \multirow{2}*{BZ107C2} & \multirow{2}*{2024.829434} & PR & 1.834$\pm$0.067 & 27.3 & \makebox[20pt][r]{37.095} & \makebox[20pt][r]{34.569} & 0.025 & 0.058 \\
         & & & sMV & 2.275$\pm$0.041 & 55.4 & \makebox[20pt][r]{37.320} & \makebox[20pt][r]{34.230} & 0.011 & 0.025 \\
         \hline
         \multirow{4}*{AR Mon} & \multirow{2}*{BZ087C1} & \multirow{2}*{2021.678721} & PR & 0.276$\pm$0.028 & 9.8 & 07:20:48.4647689 & $-$05:15:35.943597 & 0.054 & 0.142 \\
         & & & sMV & 0.311$\pm$0.028 & 11.3 & 07:20:48.4647784 & $-$05:15:35.943515 & 0.049 & 0.125 \\
         & \multirow{2}*{BZ087C2} & \multirow{2}*{2021.804318} & PR & 1.503$\pm$0.045 & 33.4 & \makebox[20pt][r]{1.002} & \makebox[20pt][r]{-1.852} & 0.022 & 0.055 \\
         & & & sMV & 1.657$\pm$0.034 & 49.4 & \makebox[20pt][r]{1.253} & \makebox[20pt][r]{-1.711} & 0.015 & 0.036 \\
         \hline
         \multirow{6}*{XY UMa} & \multirow{2}*{BZ087D1} & \multirow{2}*{2021.771742} & PR & 0.212$\pm$0.031 & 6.8 & 09:09:55.8125729 & $+$54:29:13.740698 & 0.133 & 0.158 \\
         & & & sMV & 0.291$\pm$0.036 & 8.2 & 09:09:55.8125171 & $+$54:29:13.740163 & 0.133 & 0.138 \\
         & \multirow{2}*{BZ087D2} & \multirow{2}*{2021.930104} & PR & 0.474$\pm$0.035 & 13.7 & \makebox[20pt][r]{-8.208} & \makebox[20pt][r]{-20.022} & 0.107 & 0.112 \\
         & & & sMV & 0.474$\pm$0.035 & 13.6 & \makebox[20pt][r]{-8.784} & \makebox[20pt][r]{-20.595} & 0.109 & 0.111 \\
         & \multirow{2}*{BZ103D2} & \multirow{2}*{2024.343721} & PR & 0.187$\pm$0.033 & 5.6 & \makebox[20pt][r]{-154.854} & \makebox[20pt][r]{-459.799} & 0.333 & 0.520 \\
         & & & sMV & 0.229$\pm$0.034 & 6.7 & \makebox[20pt][r]{-155.240} & \makebox[20pt][r]{-460.258} & 0.257 & 0.376 \\
         \hline
         \multirow{14}*{FF UMa} & \multirow{2}*{BZ087D1} & \multirow{2}*{2021.771867} & PR & 1.583$\pm$0.044 & 35.8 & 09:33:46.4844440 & $+$62:49:39.766235 & 0.040 & 0.047 \\
         & & & sMV & 1.687$\pm$0.041 & 41.1 & 09:33:46.4844371 & $+$62:49:39.766465 & 0.034 & 0.039 \\
         & \multirow{2}*{BZ087D2} & \multirow{2}*{2021.930229} & PR & 1.932$\pm$0.045 & 43.0 & \makebox[20pt][r]{-2.276} & \makebox[20pt][r]{1.769} & 0.051 & 0.033 \\
         & & & sMV & 2.163$\pm$0.039 & 55.0 & \makebox[20pt][r]{-2.334} & \makebox[20pt][r]{2.169} & 0.036 & 0.025 \\
         & \multirow{2}*{BZ103D1} & \multirow{2}*{2024.283813} & PR & 7.684$\pm$0.131 & 58.5 & \makebox[20pt][r]{-63.833} & \makebox[20pt][r]{-46.719} & 0.024 & 0.019 \\
         & & & sMV & 8.039$\pm$0.105 & 76.6 & \makebox[20pt][r]{-63.878} & \makebox[20pt][r]{-46.306} & 0.018 & 0.014 \\
         & \multirow{2}*{BZ103D2} & \multirow{2}*{2024.343846} & PR & 1.244$\pm$0.052 & 24.1 & \makebox[20pt][r]{-66.210} & \makebox[20pt][r]{-49.814} & 0.086 & 0.060 \\
         & & & sMV & 1.363$\pm$0.049 & 28.0 & \makebox[20pt][r]{-66.241} & \makebox[20pt][r]{-49.301} & 0.071 & 0.048 \\
         & \multirow{2}*{BZ103D3} & \multirow{2}*{2024.425792} & PR & 2.138$\pm$0.035 & 60.6 & \makebox[20pt][r]{-67.384} & \makebox[20pt][r]{-54.893} & 0.028 & 0.021 \\
         & & & sMV & 2.186$\pm$0.034 & 65.3 & \makebox[20pt][r]{-67.572} & \makebox[20pt][r]{-54.408} & 0.026 & 0.019 \\
         & \multirow{2}*{BZ107D1} & \multirow{2}*{2024.797123} & PR & 18.223$\pm$0.317 & 57.4 & \makebox[20pt][r]{-59.918} & \makebox[20pt][r]{-66.789} & 0.037 & 0.020 \\
         & & & sMV & 20.479$\pm$0.235 & 87.1 & \makebox[20pt][r]{-59.882} & \makebox[20pt][r]{-66.397} & 0.022 & 0.013 \\
         & \multirow{2}*{BZ107D2} & \multirow{2}*{2024.933642} & PR & 0.692$\pm$0.035 & 19.6 & \makebox[20pt][r]{-62.808} & \makebox[20pt][r]{-64.809} & 0.142 & 0.073 \\
         & & & sMV & 0.782$\pm$0.034 & 22.8 & \makebox[20pt][r]{-62.874} & \makebox[20pt][r]{-64.115} & 0.110 & 0.060 \\
         \hline
         \multirow{12}*{DM UMa} & \multirow{2}*{BZ087D1} & \multirow{2}*{2021.771984} & PR & 0.576$\pm$0.035 & 16.5 & 10:55:43.4325037 & $+$60:28:09.551923 & 0.080 & 0.085 \\
         & & & sMV & 0.389$\pm$0.037 & 10.6 & 10:55:43.4323585 & $+$60:28:09.548613 & 0.155 & 0.146 \\
         & \multirow{2}*{BZ087D2} & \multirow{2}*{2021.930345} & PR & 0.688$\pm$0.042 & 16.5 & \makebox[20pt][r]{-3.660} & \makebox[20pt][r]{1.788} & 0.087 & 0.091 \\
         & & & sMV & 0.581$\pm$0.045 & 12.9 & \makebox[20pt][r]{-4.342} & \makebox[20pt][r]{0.386} & 0.111 & 0.123 \\
         & \multirow{2}*{BZ103D1} & \multirow{2}*{2024.283929} & PR & 0.566$\pm$0.043 & 13.3 & \makebox[20pt][r]{-101.085} & \makebox[20pt][r]{-10.588} & 0.119 & 0.088 \\
         & & & sMV & 0.499$\pm$0.044 & 11.2 & \makebox[20pt][r]{-101.562} & \makebox[20pt][r]{-11.738} & 0.135 & 0.110 \\
         & \multirow{2}*{BZ103D2} & \multirow{2}*{2024.343962} & PR & 1.801$\pm$0.069 & 26.2 & \makebox[20pt][r]{-104.653} & \makebox[20pt][r]{-12.142} & 0.077 & 0.060 \\
         & & & sMV & 1.057$\pm$0.102 & 10.4 & \makebox[20pt][r]{-105.393} & \makebox[20pt][r]{-13.113} & 0.205 & 0.163 \\
         & \multirow{2}*{BZ103D3} & \multirow{2}*{2024.425909} & PR & 0.318$\pm$0.031 & 10.3 & \makebox[20pt][r]{-108.324} & \makebox[20pt][r]{-14.821} & 0.209 & 0.142 \\
         & & & sMV & 0.251$\pm$0.032 & 7.7 & \makebox[20pt][r]{-108.794} & \makebox[20pt][r]{-16.031} & 0.248 & 0.215 \\
         & \multirow{2}*{BZ107D1} & \multirow{2}*{2024.797240} & PR & 0.803$\pm$0.054 & 14.9 & \makebox[20pt][r]{-113.988} & \makebox[20pt][r]{-22.986} & 0.184 & 0.099 \\
         & & & sMV & 0.719$\pm$0.060 & 12.0 & \makebox[20pt][r]{-114.694} & \makebox[20pt][r]{-24.479} & 0.213 & 0.133 \\
         \pagebreak
         \multirow{12}*{RS CVn} & \multirow{2}*{BZ087E1} & \multirow{2}*{2021.608702} & PR & 3.569$\pm$0.066 & 54.1 & 13:10:36.8185054 & $+$35:56:06.032171 & 0.026 & 0.025 \\
         & & & sMV & 3.575$\pm$0.068 & 52.8 & 13:10:36.8185116 & $+$35:56:06.031949 & 0.027 & 0.026 \\
         & \multirow{2}*{BZ087E2} & \multirow{2}*{2021.630576} & PR & 2.407$\pm$0.074 & 32.3 & \makebox[20pt][r]{-0.601} & \makebox[20pt][r]{0.416} & 0.064 & 0.044 \\
         & & & sMV & 2.480$\pm$0.072 & 34.7 & \makebox[20pt][r]{-0.803} & \makebox[20pt][r]{0.131} & 0.059 & 0.041 \\
         & \multirow{2}*{BZ103E1} & \multirow{2}*{2024.107030} & PR & 0.734$\pm$0.036 & 20.4 & \makebox[20pt][r]{-112.861} & \makebox[20pt][r]{52.828} & 0.071 & 0.064 \\
         & & & sMV & 0.737$\pm$0.036 & 20.6 & \makebox[20pt][r]{-112.963} & \makebox[20pt][r]{52.908} & 0.070 & 0.063 \\
         & \multirow{2}*{BZ103E2} & \multirow{2}*{2024.197093} & PR & 0.870$\pm$0.037 & 23.7 & \makebox[20pt][r]{-120.131} & \makebox[20pt][r]{57.650} & 0.079 & 0.070 \\
         & & & sMV & 0.877$\pm$0.037 & 23.9 & \makebox[20pt][r]{-120.259} & \makebox[20pt][r]{57.767} & 0.078 & 0.069 \\
         & \multirow{2}*{BZ107E1} & \multirow{2}*{2024.633992} & PR & 0.423$\pm$0.035 & 12.2 & \makebox[20pt][r]{-150.792} & \makebox[20pt][r]{61.924} & 0.158 & 0.099 \\
         & & & sMV & 0.428$\pm$0.035 & 12.4 & \makebox[20pt][r]{-150.844} & \makebox[20pt][r]{61.798} & 0.155 & 0.097 \\
         & \multirow{2}*{BZ107E2} & \multirow{2}*{2024.680392} & PR & 1.463$\pm$0.049 & 29.6 & \makebox[20pt][r]{-151.395} & \makebox[20pt][r]{61.470} & 0.064 & 0.041 \\
         & & & sMV & 1.450$\pm$0.050 & 28.9 & \makebox[20pt][r]{-151.505} & \makebox[20pt][r]{61.414} & 0.068 & 0.043 \\
         \hline
         \multirow{10}*{RS UMi} & \multirow{2}*{BZ087E1} & \multirow{2}*{2021.608417} & PR & 0.303$\pm$0.033 & 9.2 & 15:50:49.4485186 & $+$72:12:40.443268 & 0.345 & 0.177 \\
         & & & sMV & 0.197$\pm$0.034 & 5.8 & 15:50:49.4483801 & $+$72:12:40.443232 & 0.578 & 0.264 \\
         & \multirow{2}*{BZ087E3} & \multirow{2}*{2021.949714} & PR & 0.514$\pm$0.035 & 14.5 & \makebox[20pt][r]{4.062} & \makebox[20pt][r]{-5.358} & 0.227 & 0.097 \\
         & & & sMV & 0.520$\pm$0.035 & 14.7 & \makebox[20pt][r]{3.354} & \makebox[20pt][r]{-4.780} & 0.222 & 0.095 \\
         & \multirow{2}*{BZ103E2} & \multirow{2}*{2024.196808} & PR & 1.031$\pm$0.048 & 21.3 & \makebox[20pt][r]{12.644} & \makebox[20pt][r]{-21.885} & 0.144 & 0.077 \\
         & & & sMV & 1.197$\pm$0.042 & 28.8 & \makebox[20pt][r]{12.044} & \makebox[20pt][r]{-21.153} & 0.111 & 0.049 \\
         & \multirow{2}*{BZ107E1} & \multirow{2}*{2024.633709} & PR & 1.921$\pm$0.047 & 41.0 & \makebox[20pt][r]{9.827} & \makebox[20pt][r]{-26.122} & 0.073 & 0.037 \\
         & & & sMV & 1.677$\pm$0.041 & 41.3 & \makebox[20pt][r]{9.399} & \makebox[20pt][r]{-26.698} & 0.076 & 0.039 \\
         & \multirow{2}*{BZ107E2} & \multirow{2}*{2024.680108} & PR & 2.029$\pm$0.043 & 46.6 & \makebox[20pt][r]{10.209} & \makebox[20pt][r]{-26.927} & 0.059 & 0.032 \\
         & & & sMV & 1.373$\pm$0.043 & 32.2 & \makebox[20pt][r]{9.583} & \makebox[20pt][r]{-27.193} & 0.094 & 0.053 \\
    \end{longtable}
    \begin{tablenotes}
        \makebox[0.95\textwidth]{%
            \footnotesize
            \begin{minipage}{0.95\textwidth}
                \item All results are produced by the AIPS task $\texttt{JMFIT}$.
                Column ``Tech.'' denotes calibration technique, PR or sMV.
                $S_{\mathrm{peak}}$ denotes peak flux density, and SNR denotes signal-to-noise ratio in image.
                The coordinates in the first two rows of each star are taken as reference positions for PR and sMV, respectively, and the offsets in other rows are relative to them.
            \end{minipage}
        }
    \end{tablenotes}
\end{ThreePartTable}

\twocolumn

\section{Estimated parallax curves of radio stars}
\label{app:curve}

The estimated parallax curves for ten stars are presented here, calibrated using both PR and sMV.

\begin{figure*}
    \centering
    \includegraphics[width=0.55\textwidth]{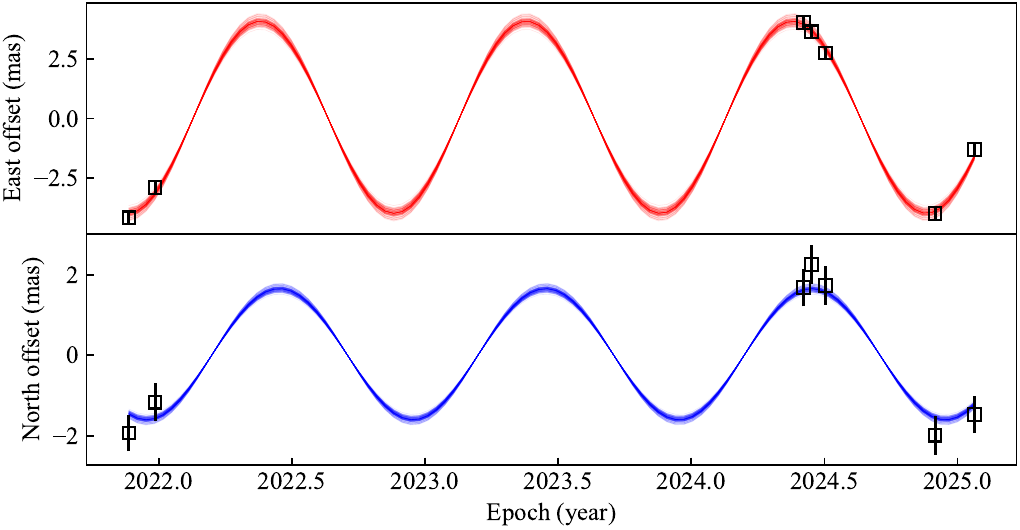}
    \includegraphics[width=0.55\textwidth]{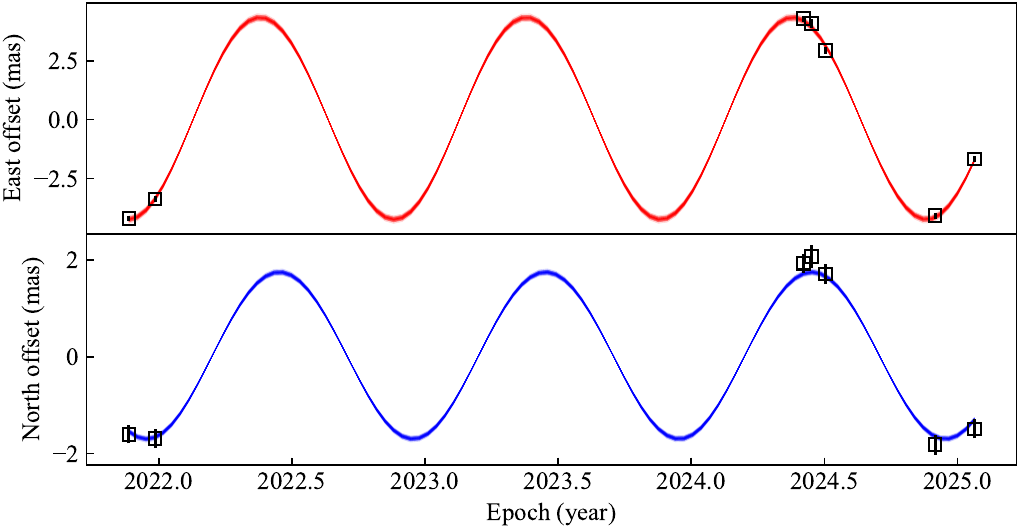}
    \caption{Estimated parallax curves of FF Aqr.
    Upper panel: calibrated with PR; Lower panel: calibrated with sMV.
    The broadening of the curves represents the probability distribution of estimated astrometric parameters.}
    \label{fig:curve_ffaqr}
\end{figure*}

\begin{figure*}
    \centering
    \includegraphics[width=0.55\textwidth]{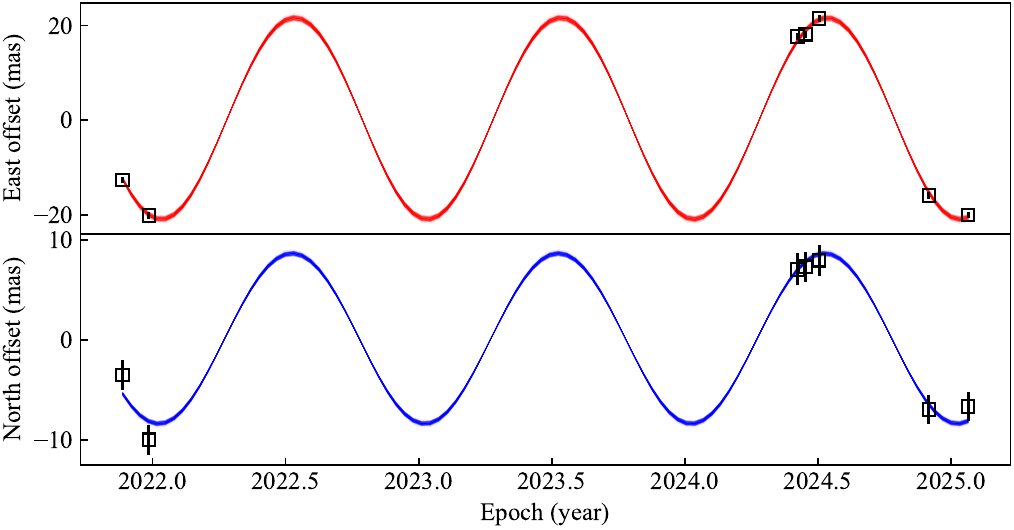}
    \includegraphics[width=0.55\textwidth]{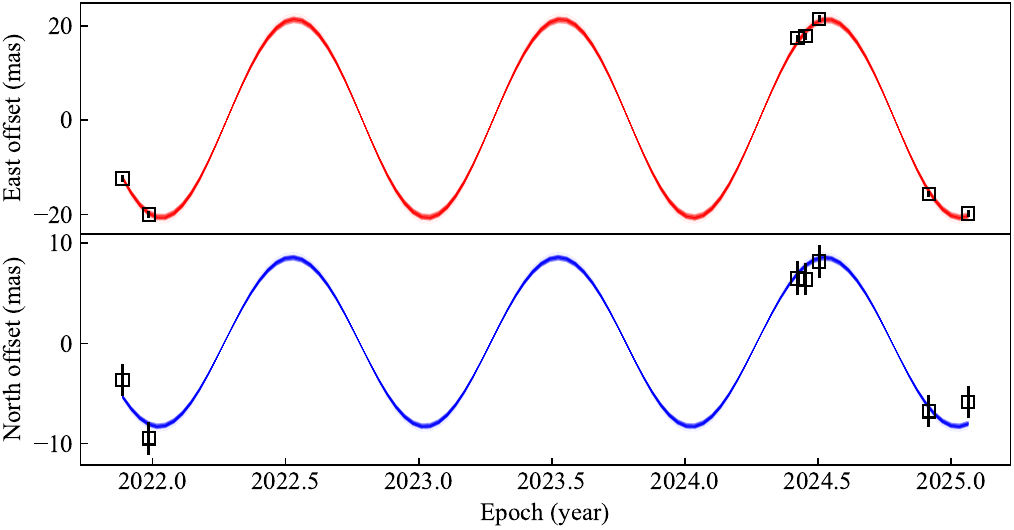}
    \caption{Estimated parallax curves of HD 8357.
    All elements in the figure are kept the same as Fig.~\ref{fig:curve_ffaqr}.}
\end{figure*}

\begin{figure*}
    \centering
    \includegraphics[width=0.55\textwidth]{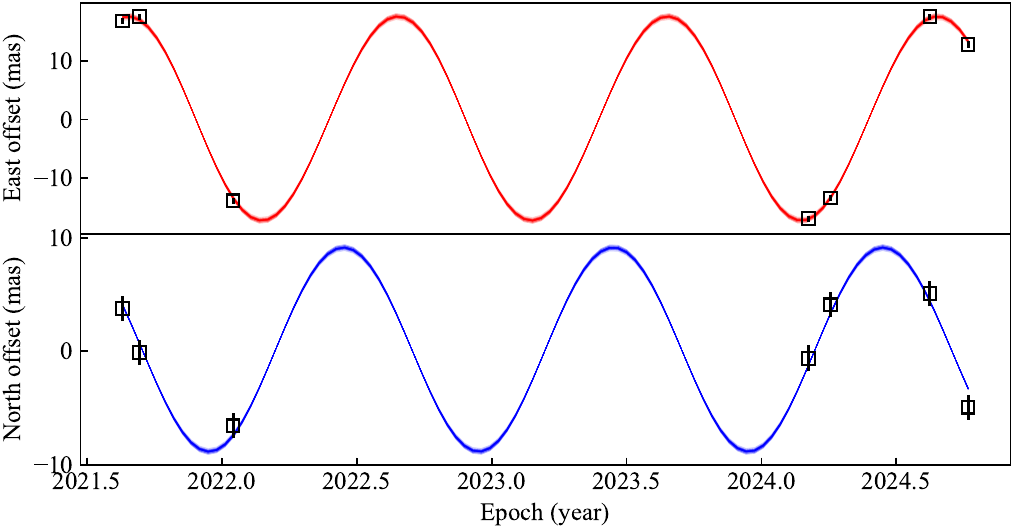}
    \includegraphics[width=0.55\textwidth]{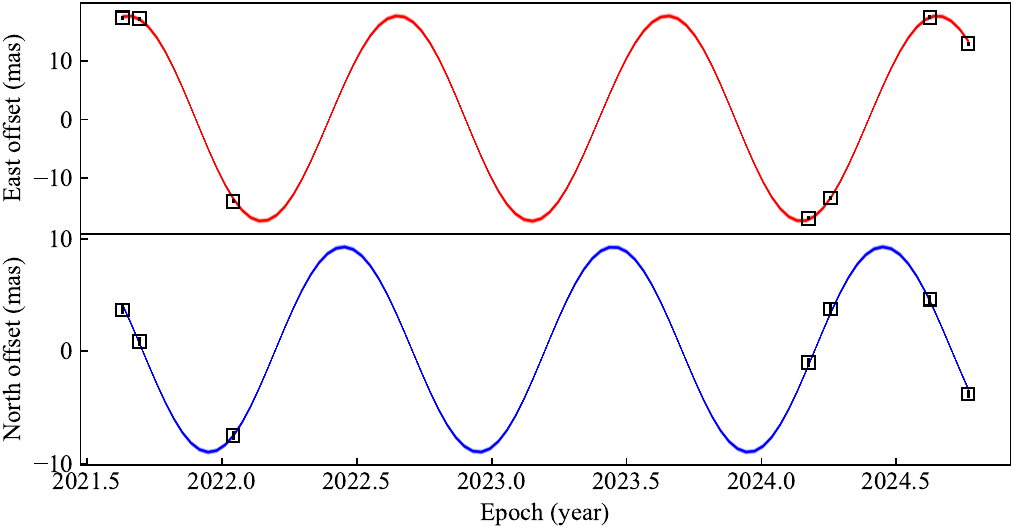}
    \caption{Estimated parallax curves of EI Eri.
    All elements in the figure are kept the same as Fig.~\ref{fig:curve_ffaqr}.}
\end{figure*}

\begin{figure*}
    \centering
    \includegraphics[width=0.55\textwidth]{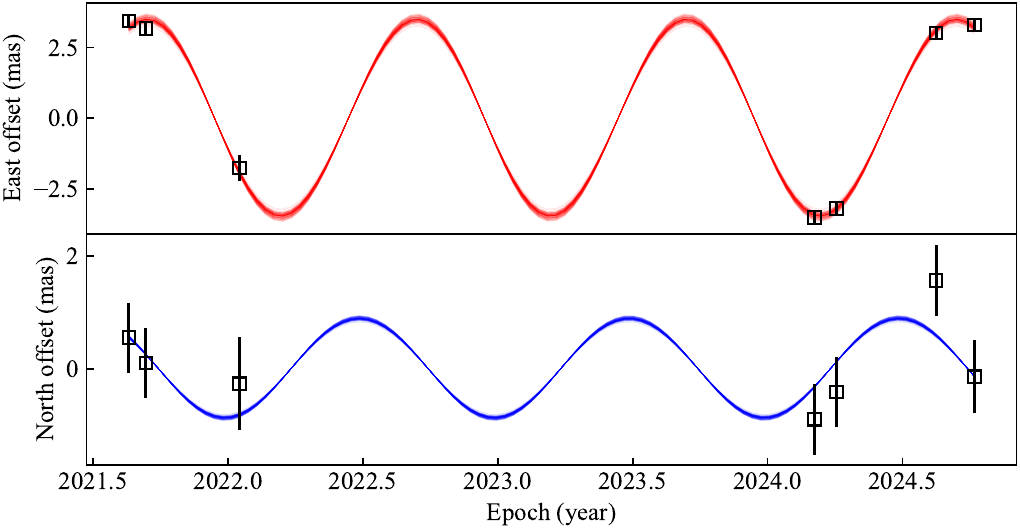}
    \includegraphics[width=0.55\textwidth]{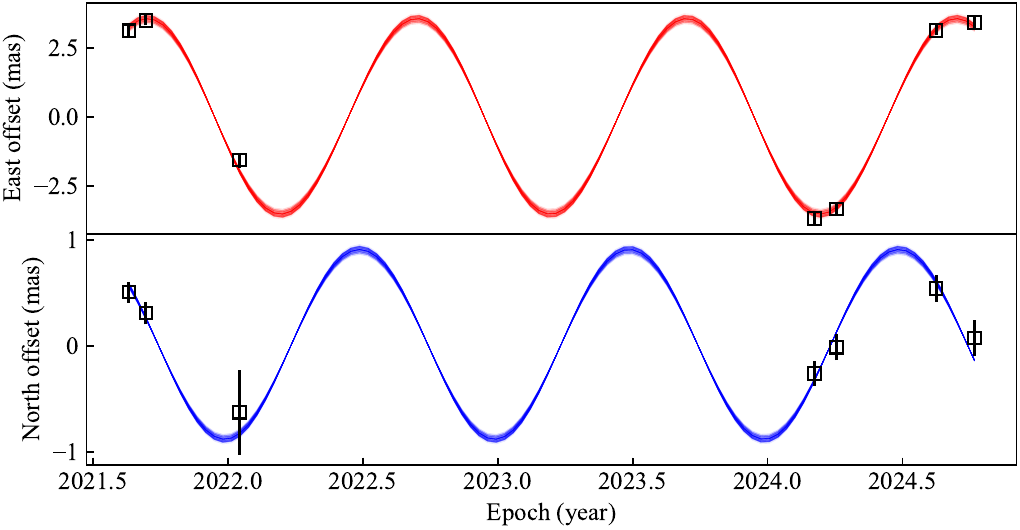}
    \caption{Estimated parallax curves of V1859 Ori.
    All elements in the figure are kept the same as Fig.~\ref{fig:curve_ffaqr}.}
\end{figure*}

\begin{figure*}
    \centering
    \includegraphics[width=0.55\textwidth]{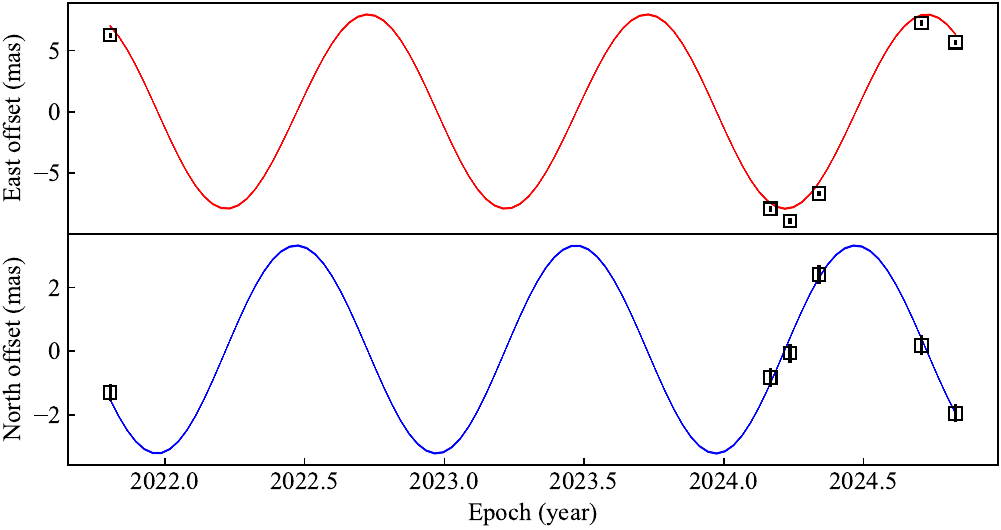}
    \includegraphics[width=0.55\textwidth]{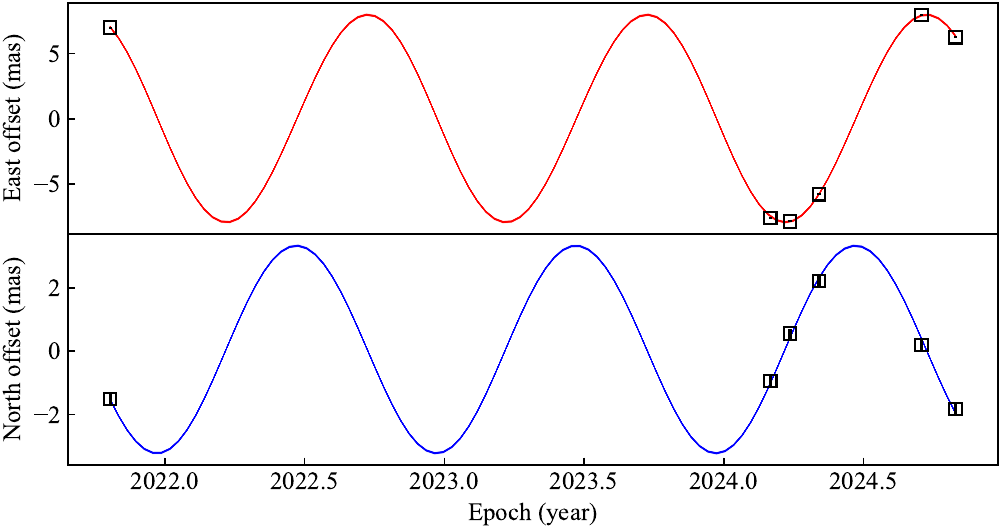}
    \caption{Estimated parallax curves of V1355 Ori.
    All elements in the figure are kept the same as Fig.~\ref{fig:curve_ffaqr}.}
\end{figure*}

\begin{figure*}
    \centering
    \includegraphics[width=0.55\textwidth]{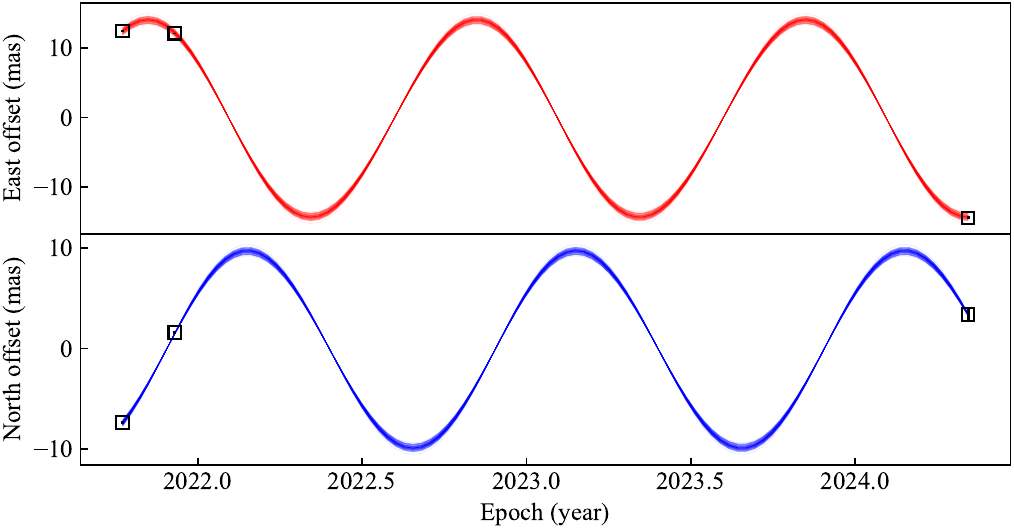}
    \includegraphics[width=0.55\textwidth]{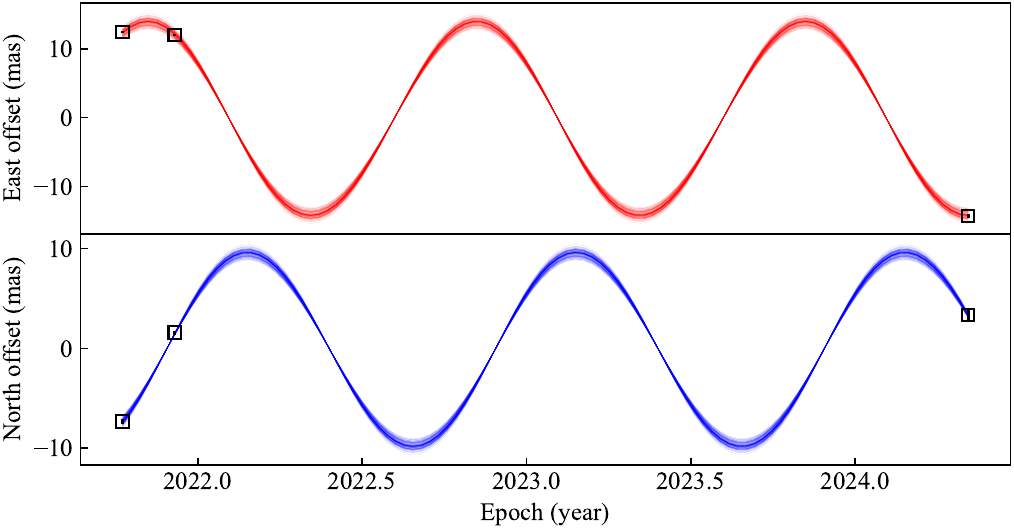}
    \caption{Estimated parallax curves of XY UMa.
    All elements in the figure are kept the same as Fig.~\ref{fig:curve_ffaqr}.}
\end{figure*}

\begin{figure*}
    \centering
    \includegraphics[width=0.55\textwidth]{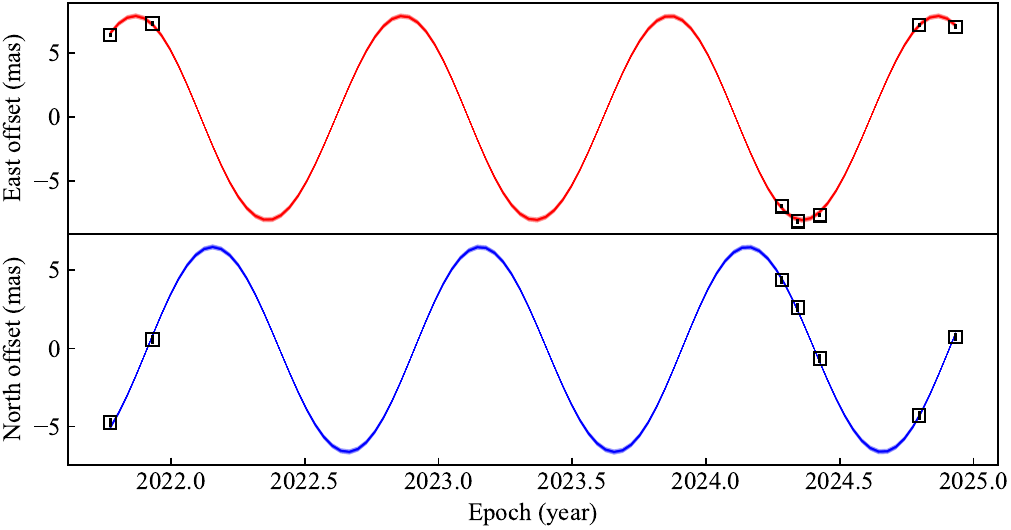}
    \includegraphics[width=0.55\textwidth]{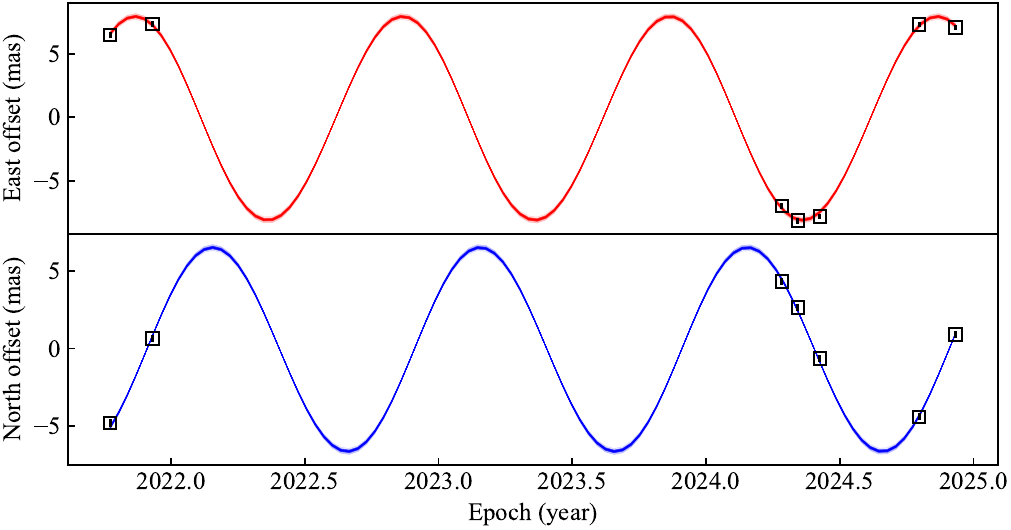}
    \caption{Estimated parallax curves of FF UMa.
    All elements in the figure are kept the same as Fig.~\ref{fig:curve_ffaqr}.}
\end{figure*}

\begin{figure*}
    \centering
    \includegraphics[width=0.55\textwidth]{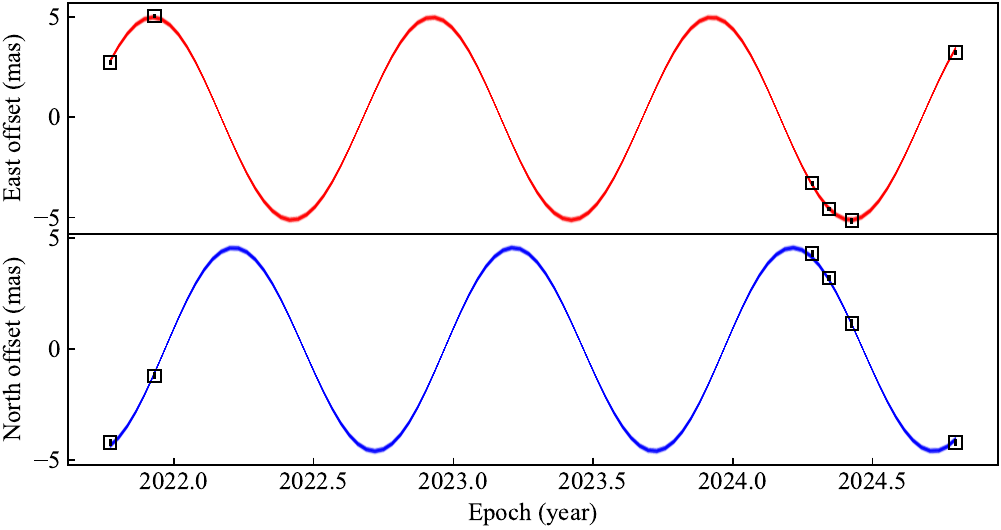}
    \includegraphics[width=0.55\textwidth]{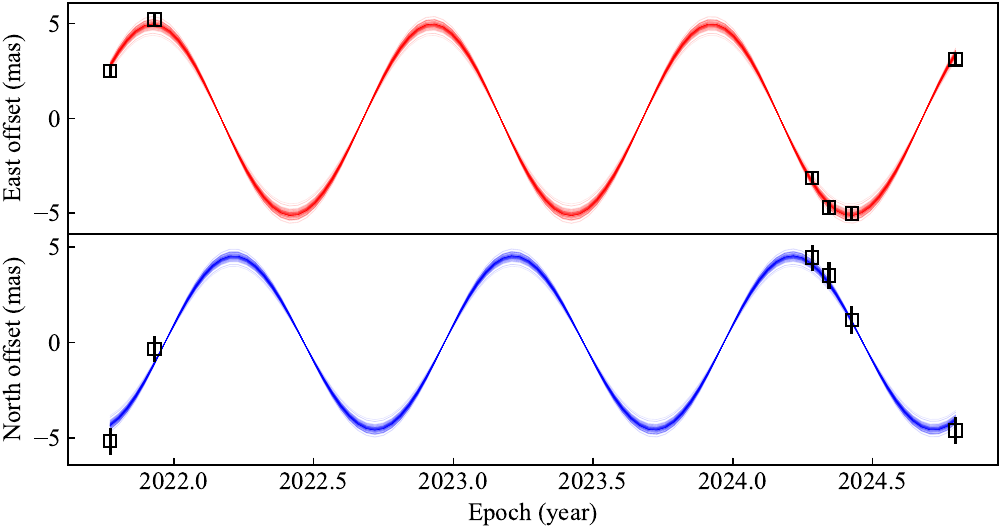}
    \caption{Estimated parallax curves of DM UMa.
    All elements in the figure are kept the same as Fig.~\ref{fig:curve_ffaqr}.}
\end{figure*}

\begin{figure*}
    \centering
    \includegraphics[width=0.55\textwidth]{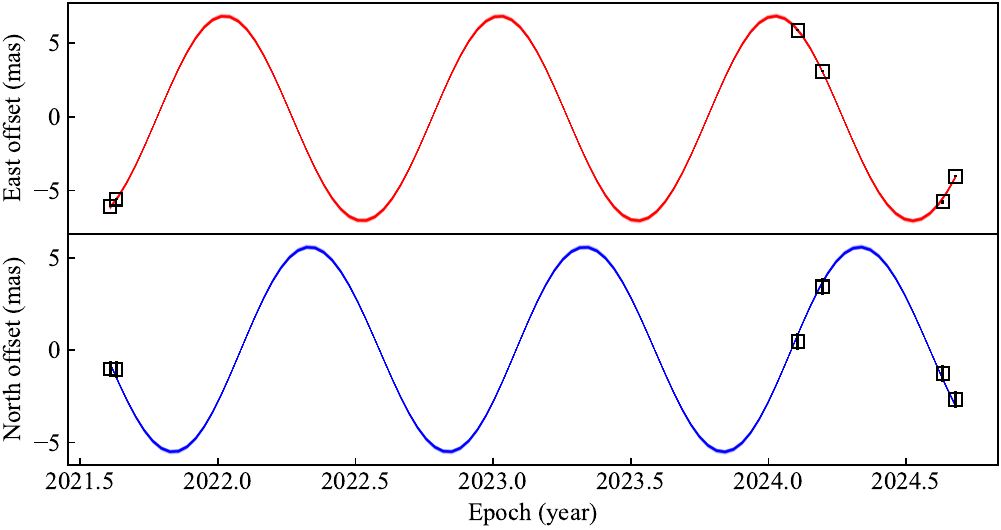}
    \includegraphics[width=0.55\textwidth]{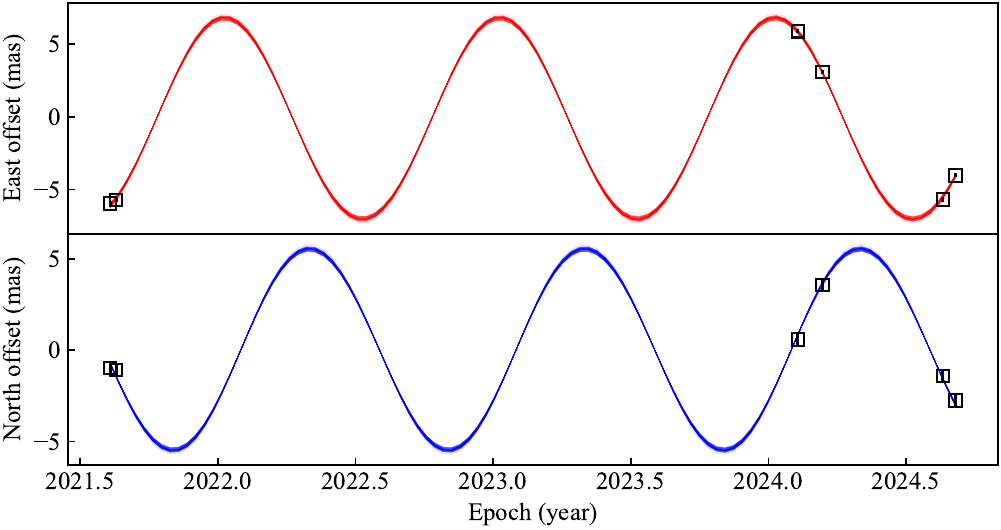}
    \caption{Estimated parallax curves of RS CVn.
    All elements in the figure are kept the same as Fig.~\ref{fig:curve_ffaqr}.}
\end{figure*}

\begin{figure*}
    \centering
    \includegraphics[width=0.55\textwidth]{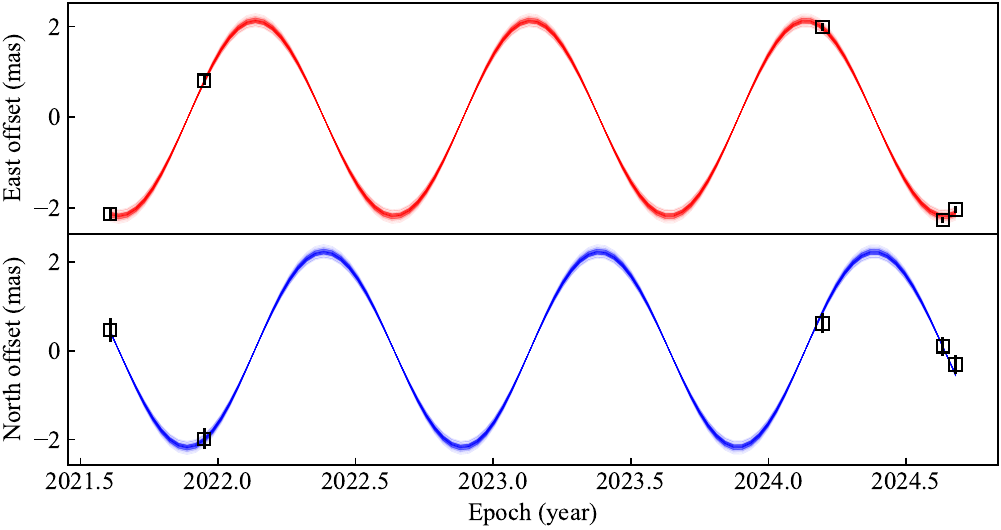}
    \includegraphics[width=0.55\textwidth]{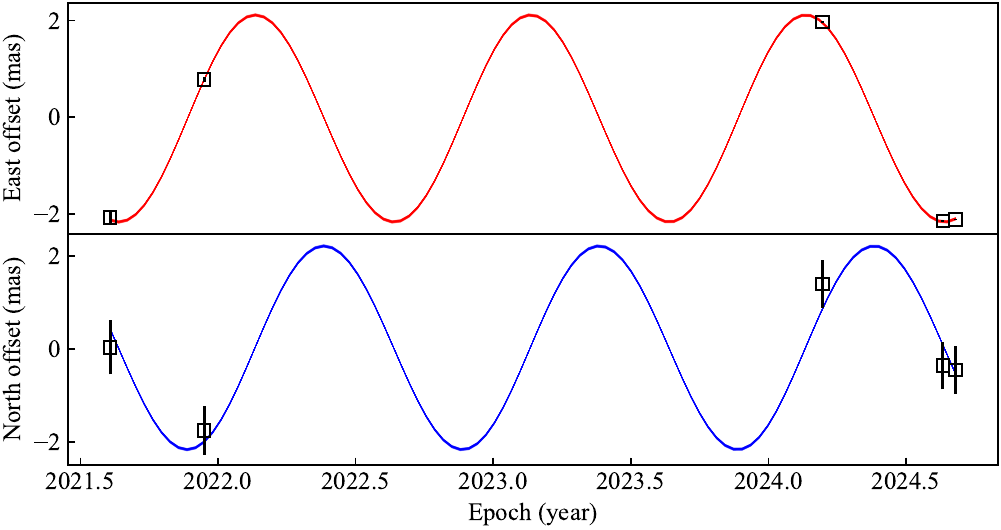}
    \caption{Estimated parallax curves of RS UMi.
    All elements in the figure are kept the same as Fig.~\ref{fig:curve_ffaqr}.}
\end{figure*}

\section{Correlation coefficient matrices of adopted astrometric solutions}
\label{app:corr}

The correlation coefficients among the astrometric parameters, $\mathrm{corr}[\alpha, \delta, \varpi, \mu_{\alpha}, \mu_{\delta}]$ (RA, DEC, parallax, and proper motions in the RA and DEC directions), are listed below.
In general, the most significant correlations occur between $\alpha$ and $\mu_{\alpha}$, and between $\delta$ and $\mu_{\delta}$, which is physically expected given the coupling between source position and linear proper motion.
A relatively high correlation is also present between $\varpi$ and the two RA-direction parameters ($\alpha$ and $\mu_{\alpha}$), because we were primarily sampling at the extremes of parallax sinusoids projected onto the RA direction, as shown in Appendix \ref{app:curve}.
The remaining terms in the correlation coefficient matrices are relatively small.

\subsubsection*{FF Aqr}
\begin{multline}
    \mathrm{corr}[\alpha, \delta, \varpi, \mu_{\alpha}, \mu_{\delta}]=\\\begin{bmatrix}
        +1.000 & +0.059 & +0.378 & -0.511 & -0.085 \\
\cdots & +1.000 & +0.112 & -0.052 & -0.413 \\
\cdots & \cdots & +1.000 & -0.425 & -0.138 \\
\cdots & \cdots & \cdots & +1.000 & +0.063 \\
\cdots & \cdots & \cdots & \cdots & +1.000
    \end{bmatrix}
\end{multline}

\subsubsection*{HD 8357}
\begin{multline}
    \mathrm{corr}[\alpha, \delta, \varpi, \mu_{\alpha}, \mu_{\delta}]=\\\begin{bmatrix}
        +1.000 & +0.012 & +0.214 & -0.509 & -0.000 \\
\cdots & +1.000 & +0.032 & -0.014 & -0.466 \\
\cdots & \cdots & +1.000 & -0.318 & -0.054 \\
\cdots & \cdots & \cdots & +1.000 & -0.008 \\
\cdots & \cdots & \cdots & \cdots & +1.000
    \end{bmatrix}
\end{multline}

\subsubsection*{EI Eri}
\begin{multline}
    \mathrm{corr}[\alpha, \delta, \varpi, \mu_{\alpha}, \mu_{\delta}]=\\\begin{bmatrix}
        +1.000 & -0.008 & -0.246 & -0.114 & +0.043 \\
\cdots & +1.000 & +0.042 & +0.001 & -0.061 \\
\cdots & \cdots & +1.000 & +0.190 & -0.023 \\
\cdots & \cdots & \cdots & +1.000 & +0.006 \\
\cdots & \cdots & \cdots & \cdots & +1.000
    \end{bmatrix}
\end{multline}

\subsubsection*{V1859 Ori}
\begin{multline}
    \mathrm{corr}[\alpha, \delta, \varpi, \mu_{\alpha}, \mu_{\delta}]=\\\begin{bmatrix}
        +1.000 & +0.041 & -0.319 & -0.204 & -0.035 \\
\cdots & +1.000 & -0.111 & -0.038 & +0.033 \\
\cdots & \cdots & +1.000 & +0.272 & +0.065 \\
\cdots & \cdots & \cdots & +1.000 & -0.003 \\
\cdots & \cdots & \cdots & \cdots & +1.000
    \end{bmatrix}
\end{multline}

\subsubsection*{V1355 Ori}
\begin{multline}
    \mathrm{corr}[\alpha, \delta, \varpi, \mu_{\alpha}, \mu_{\delta}]=\\\begin{bmatrix}
        +1.000 & -0.041 & -0.087 & -0.570 & +0.041 \\
\cdots & +1.000 & +0.042 & +0.015 & -0.592 \\
\cdots & \cdots & +1.000 & +0.205 & -0.030 \\
\cdots & \cdots & \cdots & +1.000 & -0.020 \\
\cdots & \cdots & \cdots & \cdots & +1.000
    \end{bmatrix}
\end{multline}

\subsubsection*{XY UMa}

Since XY UMa was only detected in three epochs, the correlation coefficients between astrometric parameters are extremely high.
\begin{multline}
    \mathrm{corr}[\alpha, \delta, \varpi, \mu_{\alpha}, \mu_{\delta}]=\\\begin{bmatrix}
        +1.000 & -0.107 & +0.492 & +0.683 & -0.187 \\
\cdots & +1.000 & -0.226 & -0.218 & +0.917 \\
\cdots & \cdots & +1.000 & +0.957 & -0.384 \\
\cdots & \cdots & \cdots & +1.000 & -0.369 \\
\cdots & \cdots & \cdots & \cdots & +1.000
    \end{bmatrix}
\end{multline}

\subsubsection*{FF UMa}
\begin{multline}
    \mathrm{corr}[\alpha, \delta, \varpi, \mu_{\alpha}, \mu_{\delta}]=\\\begin{bmatrix}
        +1.000 & -0.019 & -0.227 & -0.440 & -0.035 \\
\cdots & +1.000 & +0.078 & +0.014 & -0.384 \\
\cdots & \cdots & +1.000 & +0.385 & -0.123 \\
\cdots & \cdots & \cdots & +1.000 & -0.025 \\
\cdots & \cdots & \cdots & \cdots & +1.000
    \end{bmatrix}
\end{multline}

\subsubsection*{DM UMa}
\begin{multline}
    \mathrm{corr}[\alpha, \delta, \varpi, \mu_{\alpha}, \mu_{\delta}]=\\\begin{bmatrix}
        +1.000 & +0.014 & -0.077 & -0.144 & -0.013 \\
\cdots & +1.000 & +0.069 & +0.046 & -0.255 \\
\cdots & \cdots & +1.000 & +0.616 & -0.271 \\
\cdots & \cdots & \cdots & +1.000 & -0.152 \\
\cdots & \cdots & \cdots & \cdots & +1.000
    \end{bmatrix}
\end{multline}

\subsubsection*{RS CVn}
\begin{multline}
    \mathrm{corr}[\alpha, \delta, \varpi, \mu_{\alpha}, \mu_{\delta}]=\\\begin{bmatrix}
        +1.000 & -0.004 & +0.466 & -0.293 & +0.004 \\
\cdots & +1.000 & +0.015 & +0.005 & -0.151 \\
\cdots & \cdots & +1.000 & -0.539 & -0.030 \\
\cdots & \cdots & \cdots & +1.000 & +0.024 \\
\cdots & \cdots & \cdots & \cdots & +1.000
    \end{bmatrix}
\end{multline}

\subsubsection*{RS UMi}
\begin{multline}
    \mathrm{corr}[\alpha, \delta, \varpi, \mu_{\alpha}, \mu_{\delta}]=\\\begin{bmatrix}
        +1.000 & -0.033 & -0.080 & -0.813 & +0.007 \\
\cdots & +1.000 & +0.066 & +0.052 & -0.208 \\
\cdots & \cdots & +1.000 & +0.460 & -0.008 \\
\cdots & \cdots & \cdots & +1.000 & -0.017 \\
\cdots & \cdots & \cdots & \cdots & +1.000
    \end{bmatrix}
\end{multline}

\section{Self-calibrated images of calibrators}
\label{app:structure}

We present self-calibrated images of calibrators that have significant source structures here.
Epochs in which all ten VLBA antennas participated are selected for the best UV coverage.
Note that each image has a different field of view (FOV).
The cleaning and self-calibration are done using Difmap \citep{1997ASPC..125...77S}.

\begin{figure*}
    \centering
    \includegraphics[width=0.45\textwidth]{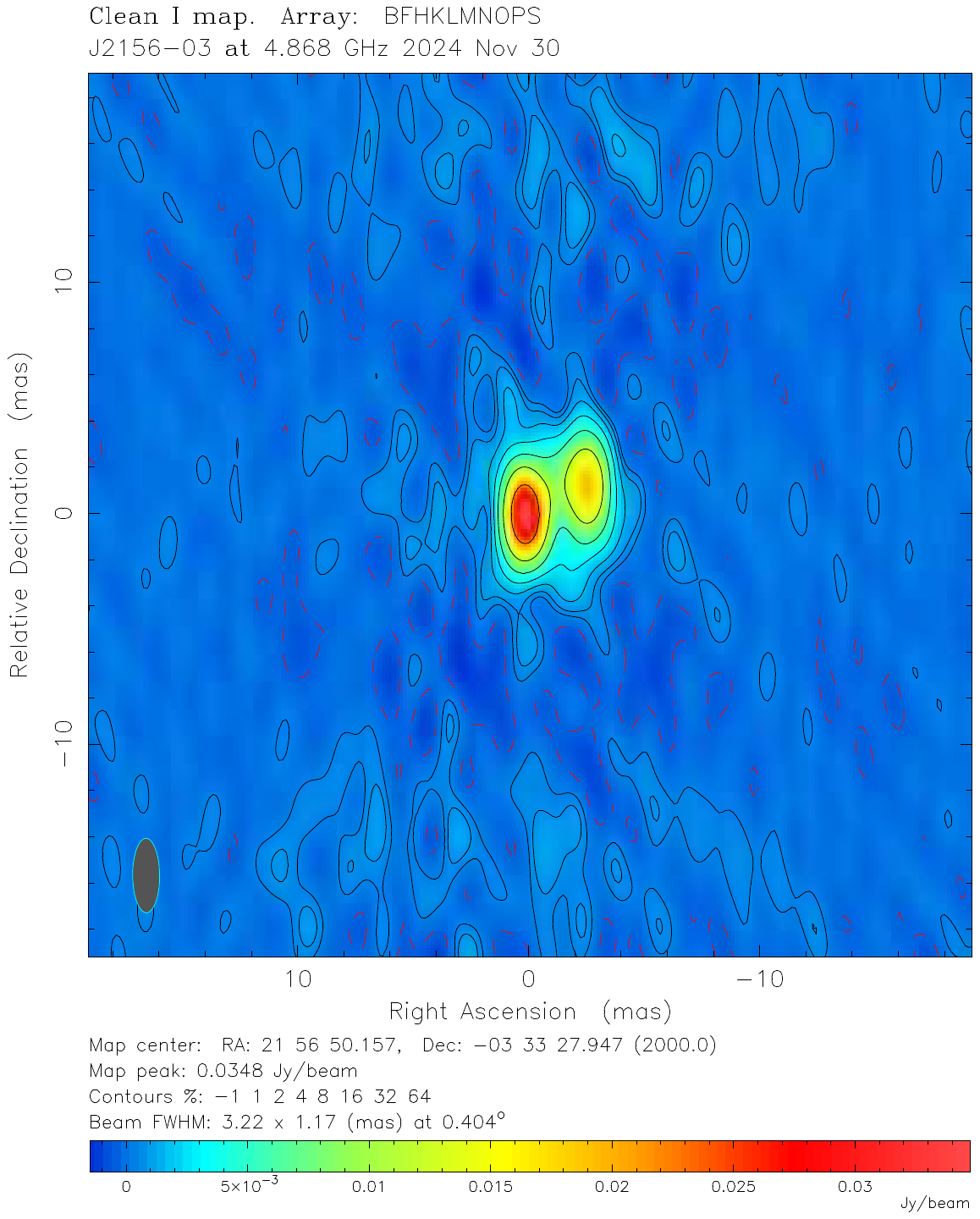}
    \includegraphics[width=0.45\textwidth]{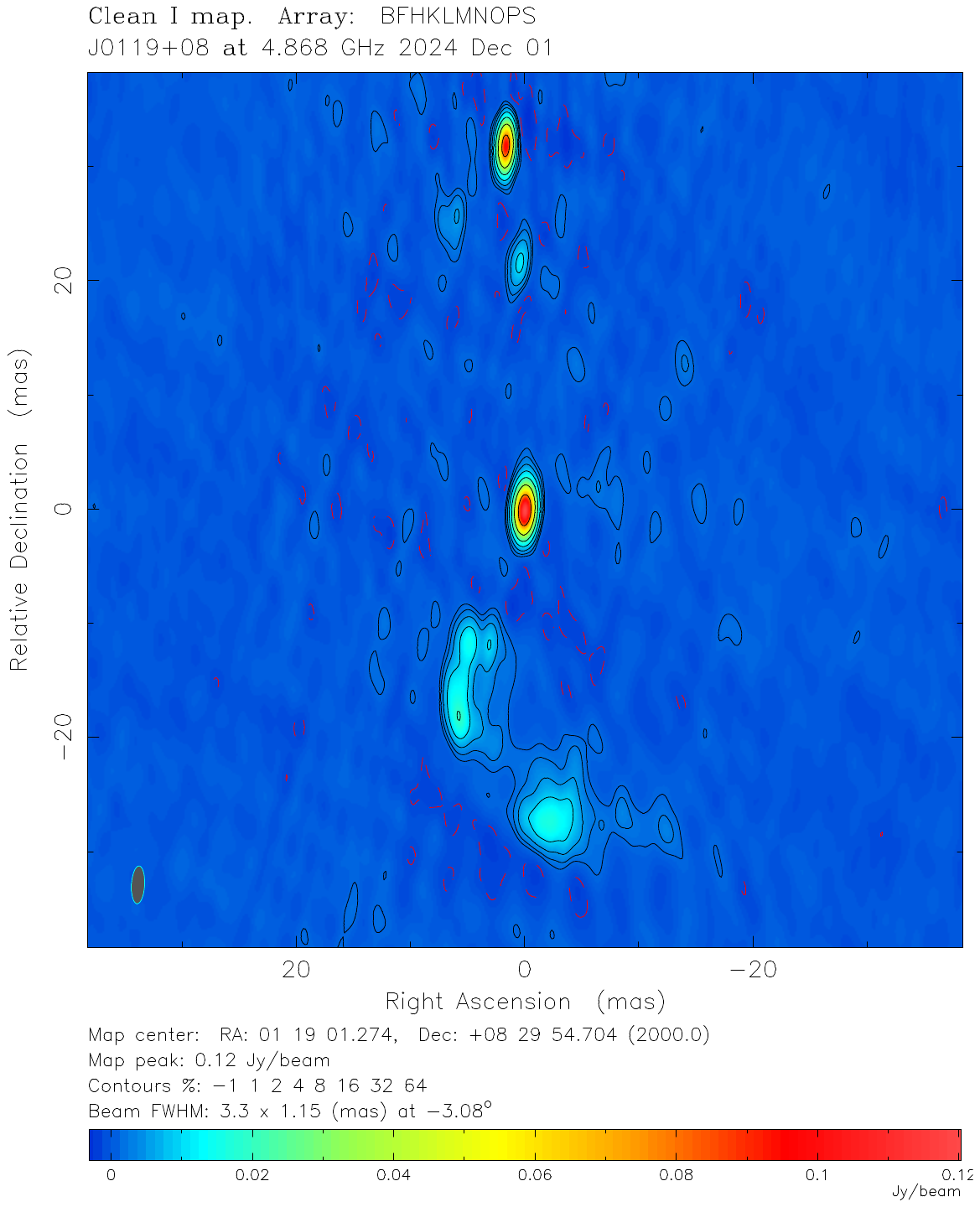}
    \includegraphics[width=0.45\textwidth]{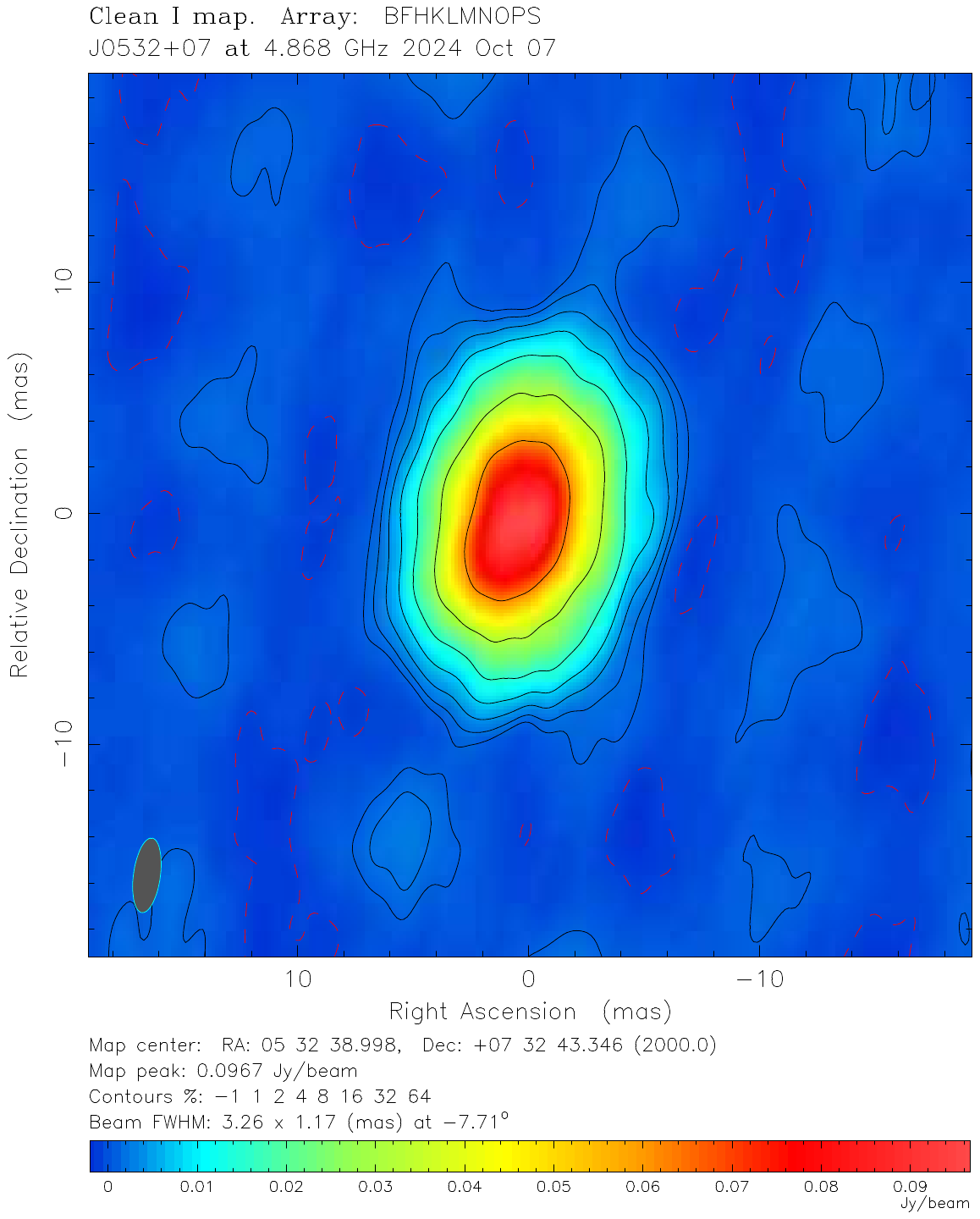}
    \includegraphics[width=0.45\textwidth]{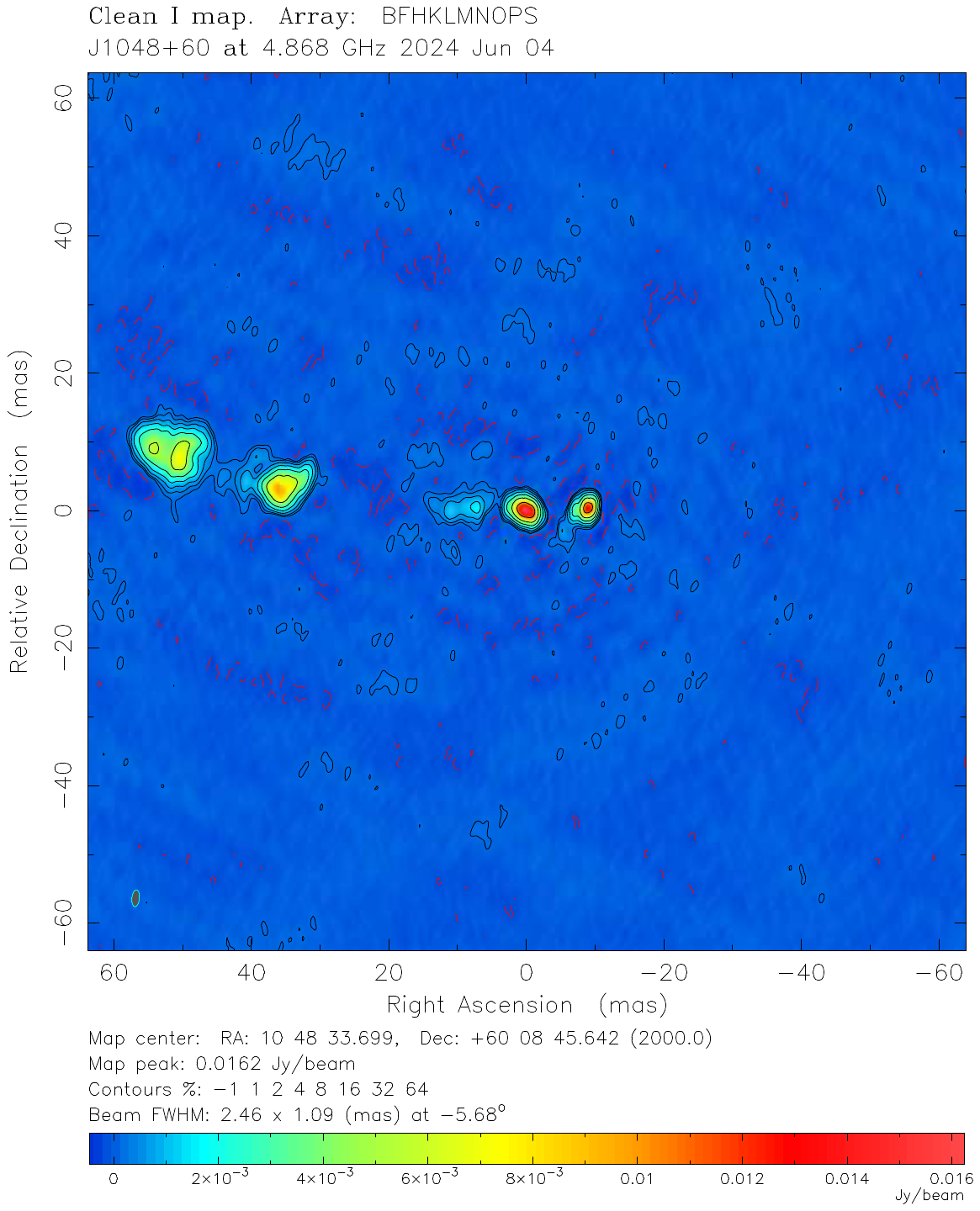}
    \caption{Self-calibrated images of: J2156-0333 (upper left); J0119+0829 (upper right); J0532+0732 (lower left); J1048+6008 (lower right).}
\end{figure*}

\begin{figure*}
    \centering
    \includegraphics[width=0.45\textwidth]{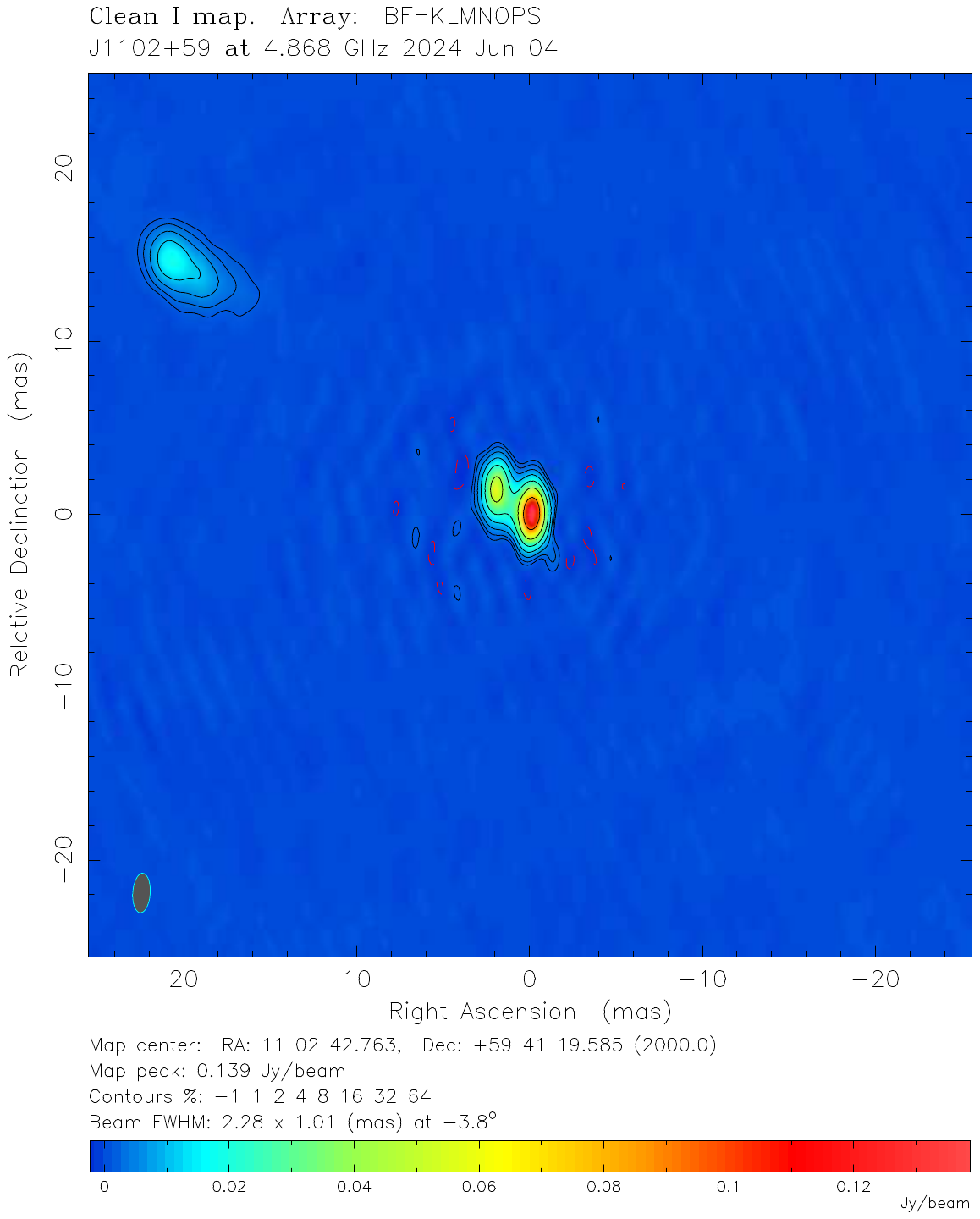}
    \includegraphics[width=0.45\textwidth]{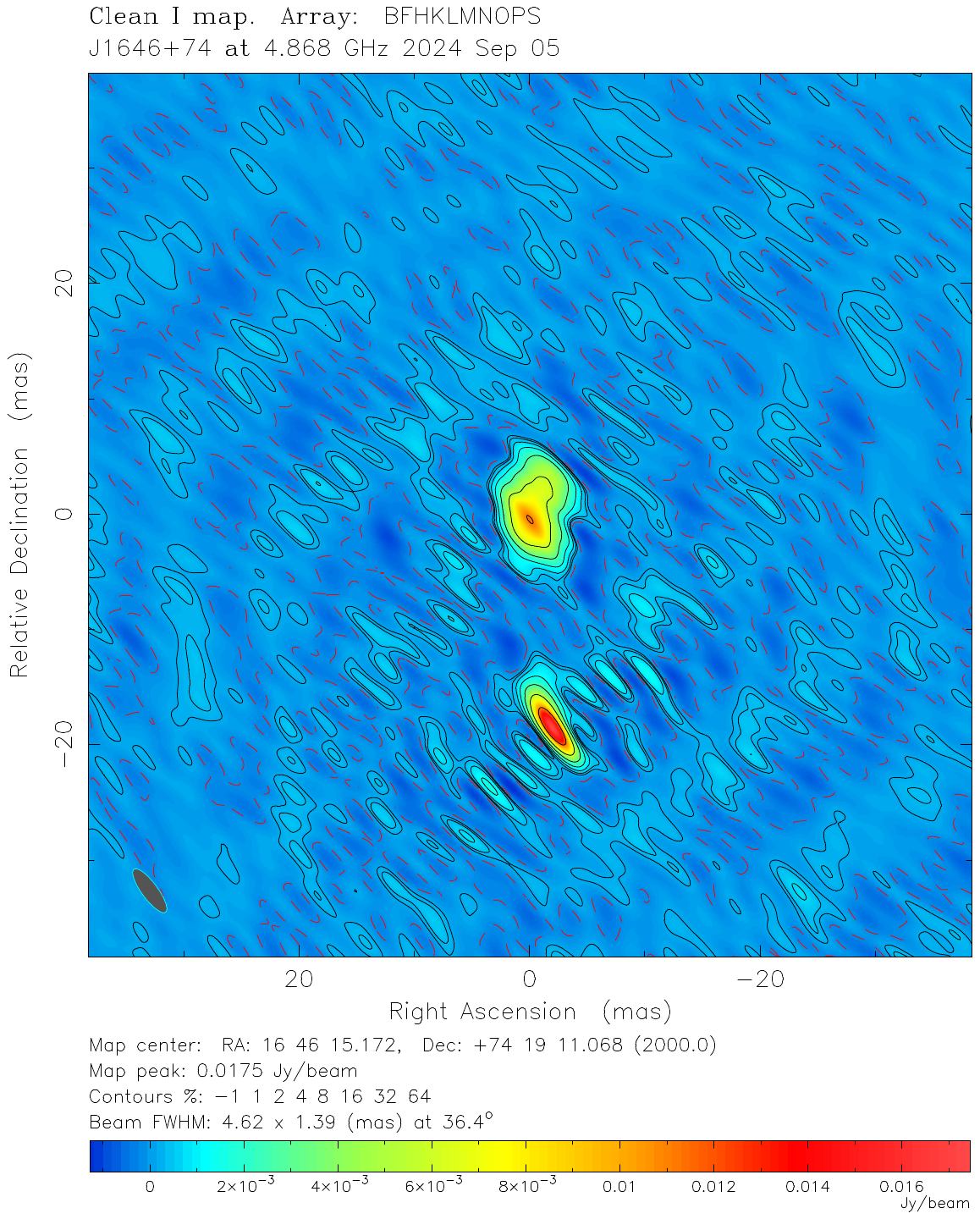}
    \caption{Self-calibrated images of: J1102+5941 (left); J1646+7419 (right).}
\end{figure*}

\bsp	
\label{lastpage}
\end{document}